\documentclass[twocolumn,showpacs,preprintnumbers,amsmath,
               nofootinbib,aps]{revtex4}

\usepackage{graphicx}
\usepackage{dcolumn}
\usepackage{bm}

\newcommand*{\bea}{\begin{eqnarray}}
\newcommand*{\eea}{\end{eqnarray}}
\newcommand*{\be}{\begin{equation}}
\newcommand*{\ee}{\end{equation}}

\newcommand*{\pref}[1]{(\ref{#1})}

\newcommand*{\mn}{{\mu\nu}}

\newcommand*{\prefr}[2]{(\ref{#1}--\ref{#2})} 
\newcommand*{\nn}{\nonumber}

\newcommand*{\1}{1\!\!\!\bot}

\begin{document}

\preprint{}

\title{Infrared properties of propagators in Landau-gauge 
       pure Yang-Mills theory at finite temperature}

\author{Attilio Cucchieri}\email{attilio@ifsc.usp.br}
\affiliation{Instituto de F\'\i sica de S\~ao Carlos, Universidade de S\~ao Paulo, \\
              Caixa Postal 369, 13560-970 S\~ao Carlos, SP, Brazil}

\author{Axel Maas\footnote{Present address: Department of Complex Physical Systems, 
Institute of Physics, Slovak Academy of Sciences, D\'{u}bravsk\'{a} cesta 9, 
SK-845 11 Bratislava, Slovakia}}\email{axel.maas@savba.sk}
\affiliation{Instituto de F\'\i sica de S\~ao Carlos, Universidade de S\~ao Paulo, \\
              Caixa Postal 369, 13560-970 S\~ao Carlos, SP, Brazil}

\author{Tereza Mendes}\email{mendes@ifsc.usp.br}
\affiliation{Instituto de F\'\i sica de S\~ao Carlos, Universidade de S\~ao Paulo, \\
              Caixa Postal 369, 13560-970 S\~ao Carlos, SP, Brazil}

\date{\today}

\begin{abstract}
The finite-temperature behavior of gluon and of 
Faddeev-Popov-ghost propagators is investigated for pure $SU(2)$ Yang-Mills 
theory in Landau gauge. We present nonperturbative results, 
obtained using lattice simulations and Dyson-Schwinger equations. 
Possible limitations of these two approaches, such as finite-volume 
effects and truncation artifacts, are extensively discussed.
Both methods suggest a very different temperature dependence for the magnetic sector
when compared to the electric one. In particular, a clear thermodynamic transition seems
to affect only the electric sector.  These results imply in particular the confinement of
transverse gluons at all temperatures and they can be understood inside the 
framework of the so-called Gribov-Zwanziger scenario of confinement.
\end{abstract}

\maketitle


\section{Introduction}

Thermodynamic observables, such as the free energy, show a thermodynamic
transition in Yang-Mills theory at a finite temperature $T_c$ \cite{Karsch:2001cy,Karsch:2003jg}.
This phase transition separates a low-temperature phase --- which is expected to be
highly non-perturbative and characterized by quark and gluon confinement --- from a
(in principle) perturbative high-temperature phase, where color charges should be deconfined.
Indeed, due to asymptotic freedom, the running coupling constant $g(T)$ vanishes
with increasing temperature \cite{LeBellac} and the quark-gluon plasma can be viewed
as a weakly-coupled system. 
At the same time, across the phase transition, the long-distance potential between fundamental
test-charges changes from a linearly-rising confining potential to an essentially flat one
\cite{Kaczmarek:2004gv}, i.e.\ the (temporal) string tension is zero at high temperature.

On the other hand, non-perturbative phenomena have to be expected
even for an arbitrarily small coupling at the renormalization scale \cite{Haag:1992hx}.
In particular, several studies have already pointed out that non-perturbative effects
should be present in Yang-Mills theory at any temperature, i.e.\ the high-temperature phase
should also be highly non-trivial.
For example, it is well-known that the magnetic sector is affected by strong
non-perturbative infrared (IR) problems \cite{Linde:1980ts,Maas:2005ym}.
Moreover, at finite temperature, the spatial string tension is non-vanishing \cite{Bali}.
Finally, the formal infinite-temperature limit of the theory, i.e.\ the 3-dimensional
reduced theory \cite{Maas:2005ym,Appelquist:1981vg}, is still a confining theory
\cite{Maas:2005ym,3d,Zwanziger,Cucchieri:2004mf,Maas:2004se}.
Thus, a clear description of Yang-Mills theory as a function of the temperature $T$ 
is still lacking and the fate of confinement at large T is still an open question.
In particular, one should reconcile the confining properties of the dimensionally-reduced theory with
the vanishing of the conventional string tension at large temperature.

In order to make contact with the continuum formulation of $SU(N_c)$ Yang-Mills theory, it is
necessary to consider quantities living either in the $su(N_c)$ algebra or in the continuum gauge
group $SU(N_c)/Z_N$ \cite{O'Raifeartaigh:1986vq}. 
This clearly limits the usefulness of center observables such as the Polyakov line, 
often employed in lattice simulations and in effective theories.
On the other hand, propagators of the elementary degrees of freedom of the theory, such 
as gluon and ghost fields, fulfill the above condition and provide access to non-perturbative 
aspects of the theory \cite{Alkofer:2000wg,Fischer:2006ub}. 

In Landau gauge, at zero temperature, the framework of the Gribov-Zwanziger
\cite{Zwanziger,Gribov} and of the Kugo-Ojima \cite{Kugo}
scenarios provides a basis for understanding the confinement mechanism
in Yang-Mills theory.
In particular, these scenarios predict that the ghost propagator should be
IR enhanced, compared to the propagator of a massless particle.
At the same time, the gluon propagator should vanish in the IR limit.
These predictions are supported by results obtained using different methods
\cite{Alkofer:2000wg,Fischer:2006ub,Zwanziger,vonSmekal,Watson:2001yv,Cucchieri:2001za,
Gies,Lerche:2002ep,Bloch:2003sk,Schleifenbaum:2004id,Alkofer,Ilgenfritz:2006gp,Ilgenfritz:2006he,
Sternbeck:2005tk,Cucchieri:2006xi}.
In addition, several calculations indicate that the predictions are also (at least partially) valid in the
high-temperature phase \cite{Maas:2005ym,Cucchieri,Zahed:1999tg,Maas:2005hs} and in
the infinite-temperature limit \cite{Maas:2005ym,Zwanziger,Cucchieri:2004mf,Maas:2004se,
Cucchieri,Schleifenbaum:2004id,Cucchieri:2006xi,
Cucchieri:1999sz,Cucchieri:2003di,Cucchieri:2006tf}.

Here we evaluate gluon and ghost propagators for the $SU(2)$ case using Landau gauge,
extending earlier numerical studies that focused on the high-temperature phase
\cite{Cucchieri} and on the infinite-temperature limit \cite{Cucchieri,Cucchieri:1999sz,
Cucchieri:2003di,Cucchieri:2006tf}.
We present nonperturbative results, obtained using lattice simulations and 
Dyson-Schwinger equations (DSEs).
In Section \ref{stensor} we present a short review of the Lorentz structure
for the Landau-gauge gluon and Faddeev-Popov-ghost propagators at finite temperature.
Results using lattice gauge theory for these two propagators at various temperatures,
both below and above the thermodynamic transition, will
be reported in Section \ref{slattice}. In that section we also discuss the
influence of finite-volume effects on the numerical results. In Section
\ref{DSEq} the same propagators are studied in continuum space-time using 
DSEs, which allows us to consider the limit of
vanishing momentum $p \to 0$.
Let us recall \cite{Zwanziger} that imposing the minimal Landau gauge condition,
i.e.\ restricting to the Gribov region, does not modify the DSEs, even though it 
provides supplementary conditions for their solution.
We believe that the simultaneous use of these two nonperturbative methods
--- lattice simulations and DSEs --- 
allows a better understanding of our results and of their physical implications.
A possible interpretation of the results from lattice and from DSEs is presented
in Section \ref{sscenario}. The main components of this interpretation
are a spatial/magnetic sector,
which is essentially unaffected by temperature and by the thermodynamic transition,
and a temporal/electric sector, which does show a sensitivity to the transition.
Let us stress that the results from DSEs suggest that the difference between the 
two sectors occurs for any non-zero temperature and not only above the phase transition. 
In particular, the Gribov-Zwanziger
and Kugo-Ojima confinement mechanisms remain qualitatively unaltered
in the magnetic sector when the temperature is turned on.
Finally, we present our conclusions in Section \ref{ssum}.
Some observations about the use of asymmetric lattices in
numerical simulations are reported in Appendix \ref{saasym}.
Analytic and numerical details of the DSE analysis, presented in Section
\ref{DSEq}, are collected in Appendices \ref{sakernels}--\ref{safull}.

Preliminary results have been presented in \cite{Maas:2006qw}.


\section{Propagators at finite temperature}\label{stensor}

At finite temperature, the Euclidean symmetry is manifestly broken by the 
thermal heat bath. As a consequence, the propagator of the gluon, which
is a vector particle, can no longer be described by a single tensor
structure.
Instead, one can consider the decomposition \cite{Kapusta:tk}
\be
D_\mn^{ab}(p) \, = \, P_\mn^T(p) \, D_{T}^{ab}(p) \, + \, P_\mn^L D_L^{ab}(p) \;,
\label{eq:decomp}
\ee 
where the two tensor structures $P_\mn^T(p)$ and $P_\mn^L(p)$ are respectively
transverse and longitudinal in the (three-dimensional) spatial sub-space.
Of course, in the full 4d space both tensors must be transverse, due
to the Landau gauge condition.
In terms of the momentum components $p_\mu$ we can write (with no summation
over repeated indices) \cite{Kapusta:tk}
\bea
P_\mn^T(p) &=& (1-\delta_{\mu0})(1-\delta_{\nu0})\left(\delta_\mn-
    \frac{p_\mu p_\nu}{\vec p^2}\right)
\label{projt}\\
P_\mn^L(p) &=& P_\mn(p) \,-\,P_\mn^T(p) \; .
\label{projl}
\eea
Clearly, the tensor $P^T_\mn(p)$ is the transverse projector in the
three-dimensional spatial sub-space, while $P^L_\mn(p)$ is the complement
of $P^T_\mn(p)$ to the four-dimensional Euclidean-invariant transverse
projector $P_\mn(p)=\delta_\mn-p_\mu p_\nu/p^2$.
At the same time, the decomposition (\ref{eq:decomp}) defines the two
independent scalar propagators $D_T^{ab}(p)$ and $D_L^{ab}(p)$,
which are respectively the coefficients of the 3d-transverse and of the
3d-longitudinal projectors.
Since one expects a different IR behavior for these two functions,
it is not useful to contract the gluon propagator with the usual 
4d-transverse projector $P_\mn(p)$, i.e.\ to consider a linear combination
of the two scalar functions $D_T^{ab}(p)$ and $D_L^{ab}(p)$.
By contracting the propagator $D_\mn^{ab}(p)$ with $P^T_\mn(p)$ or
with $P_\mn^L(p)$, inserting a unit matrix in color space and
using the relation
\be
p_0\,A_0^a\,=\,-\sum_{\mu=1}^3\,p_\mu\, A_\mu^a \; ,
\ee
valid in Landau gauge, we find (for non-zero three-momentum ${\vec p}$)
the relations\footnote{Similar considerations apply also to the gluon
propagator when studied using asymmetric lattices. In Appendix \ref{saasym}
we present results for the gluon tensor structures on a strongly asymmetric lattice.}
\bea
\!\!\!\!\!\!\!D_T(p) &=& \frac{1}{(d-2)N_g} \nn \\
     & \times &  <\sum_{\mu=1}^3 \, A_\mu^a(p) A_\mu^a(-p) 
       -\frac{p_0^2}{{\vec p\,}^2} \, A_0^a(p)A_0^a(-p)>\label{dt}  \\
\!\!\!\!\!\!\!D_L(p) &=& \frac{1}{N_g} \left(1+\frac{p_0^2}{{\vec p\,}^2}\right)
                      <A_0^a(p)A_0^a(-p)> \; .
\label{dl}
\eea
Here, $d=4$ is the space-time dimension, $N_g=3$ is the number of gluons
[i.e.\ $N_g = N^2_c-1$ in the $SU(N_c)$ case] and the summation over color indices is implied.
Clearly, for $p_0=0$ the 3d-transverse propagator $D_T(p)$ coincides with the
3d gluon propagator, while the 3d-longitudinal propagator $D_L(p)$ can be
identified with the propagator of the would-be-Higgs field in the
three-dimensionally-reduced theory. Thus, we can associate
the 3d-transverse gluon propagator to the so-called
magnetic sector and the 3d-longitudinal propagator to the electric sector.
Of course, at zero temperature the two scalar propagators coincide, i.e.\
$D_T(p) \,=\,D_L(p)\,=\,D(p)$.

\begin{table*}[t]
\caption{\label{conf}
Data of the configurations considered in our numerical simulations.
The three-dimensional case will be discussed in Section \ref{svolume}.
The {\em Setup} column will be used in the text to refer to the various numerical setups.
The value of the lattice spacing $a$ has been taken from \cite{Cucchieri:2003di} for the
three-dimensional case and from \cite{Fingberg:1992ju}
for the four-dimensional case.
{\em Sweeps} indicates the number of sweeps between
two consecutive gauge-fixed measurements. The temperature has been evaluated
using Eq. \pref{eq:T}. In the column {\em Method} we indicate a reference where
numerical details (including error determination, etc.) are discussed
for that particular set of data. With the symbols
$^+$ and $^*$ we indicate that (respectively) only the gluon or only the ghost propagator 
has been considered. Note that only the results corresponding to
the {\em Setups} 16, 17, 19, and 20 have already been published. Finally,
in the {\em Setup} 6 we found one ``exceptional'' configuration that
induced very large statistical fluctuations on the results. Thus, in this case, we
needed a much larger statistics in order to acquire the same level of accuracy
obtained in the other cases.}
\begin{tabular}{|c|c|c|c|c|c|c|c|c|}
\hline
\hline
{\em Setup} & $N^4$, $N_t\times N_s^3$ or $N^3$ &
 $\beta$ & $a^{-1}$ [GeV] & {\em Configurations} & {\em Sweeps} &
 $T$ [MeV] & $V_s^{1/3}$ [fm] & {\em Method} \cr
\hline
1 & $12^4$ & 2.3 & 1.193 & 426 & 40 & 0 & 1.98 & \cite{Cucchieri:2006tf}$^+$ \cr
\hline
2 & $16^4$ & 2.3 & 1.193 & 424 & 45 & 0 & 2.64 & \cite{Cucchieri:2006tf}$^+$ \cr
\hline
3 & $20^4$ & 2.3 & 1.193 & 405 & 50 & 0 & 3.30 & \cite{Cucchieri:2006tf}$^+$ \cr
\hline
4 & $32^4$ & 2.3 & 1.193 & 30 & 320 & 0 & 5.28 & \cite{Bloch:2003sk}$^*$ \cr
\hline
5 & $10\times 14^3$ & 2.3 & 1.193 & 401 & 40 & 119 & 2.31 & \cite{Cucchieri:2006tf} \cr
\hline
6 & $10\times 20^3$ & 2.3 & 1.193 & 1426 & 45 & 119 & 3.30 & \cite{Cucchieri:2006tf} \cr
\hline
7 & $10\times 26^3$ & 2.3 & 1.193 & 405 & 50 & 119 & 4.29 & \cite{Cucchieri:2006tf} \cr
\hline
8 & $4\times 20^3$ & 2.3 & 1.193 & 444 & 40 & 298 & 3.30 & \cite{Cucchieri:2006tf} \cr
\hline
9 & $4\times 26^3$ & 2.3 & 1.193 &  405 & 45 & 298 & 4.29 & \cite{Cucchieri:2006tf} \cr
\hline
10 & $4\times 34^3$ & 2.3 & 1.193 & 405 & 50 & 298 & 5.61 & \cite{Cucchieri:2006tf} \cr
\hline
11 & $3\times 28^3$ & 2.4 & 1.651 & 429 & 50 & 550 & 3.34 & \cite{Cucchieri:2006tf} \cr
\hline
12 & $4\times 40^3$ & 2.5 & 2.309 & 360 & 50 & 577 & 3.41 & \cite{Cucchieri:2006tf} \cr
\hline
13 & $2\times 20^3$ & 2.3 & 1.193 & 437 & 40 & 597 & 3.30 & \cite{Cucchieri:2006tf} \cr
\hline
14 & $2\times 32^3$ & 2.3 & 1.193 & 410 & 45 & 597 & 5.28 & \cite{Cucchieri:2006tf} \cr
\hline
15 & $2\times 42^3$ & 2.3 & 1.193 & 400 & 50 & 597 & 6.93 & \cite{Cucchieri:2006tf} \cr
\hline
16 & $20^3$ & 4.2 & 1.136 &  6161 & 40 & 0 & 3.47 & \cite{Cucchieri:2006tf} \cr
\hline
17 & $30^3$ & 4.2 & 1.136 &  10229 & 45 & 0 & 5.20 & \cite{Cucchieri:2006tf} \cr
\hline
18 & $40^3$ & 4.2 & 1.136 &  3306 & 50 & 0 & 6.94 & \cite{Cucchieri:2006tf} \cr
\hline
19 & $80^3$ & 4.2 & 1.136 & 200 & 200 & 0 & 13.9 & \cite{Cucchieri:2003di}$^+$,
  \cite{Cucchieri:2006za}$^*$\cr
\hline
20 & $140^3$ & 4.2 & 1.136 & 30 & 250 & 0 & 24.3 & \cite{Cucchieri:2003di}$^+$\cr
\hline
\hline
\end{tabular}
\end{table*}

As for the ghost propagator $D_G(p)$, since it is a scalar function, no additional
tensor structures arise in this case.

A second consequence of considering a theory at finite temperature is that the propagators
depend separately on the energy $p_0$ and on the spatial three-momentum $|\vec p|$.
Moreover, the energy is always discrete, i.e.\ $p_0 = 2 \pi T n$ with $n$ integer.
Of course, in studying the IR properties of the theory, we are mainly interested 
in the zero (or soft) modes $p_0=0$. The other (hard) modes ($p_0 \neq 0$)
have an effective
thermal mass of $2 \pi T n$ and seem to behave like massive particles
\cite{Maas:2005ym,Maas:2005hs,Kapusta:tk,Blaizot:2001nr}.

Let us stress that these observations apply both to the lattice and to the
continuum formulations of Yang-Mills theory. In the lattice case one also has
to consider the definition of the gluon field
\be
A_{\mu}(x) \,=\, \frac{1}{2 i} \left[ \, U_{\mu}(x) \,-\,
                U_{\mu}^{\dagger}(x)\,\right]_{traceless}
\ee
as a function of the link variables $U_{\mu}(x)$, which is based on the
expansion $ U_{\mu}(x) \,\approx\, \1 \,+\, i g_0 a A_{\mu}(x) $ .
Hence, in the high-temperature (symmetry-broken) phase, one should consider only 
configurations for which the trace of the Polyakov loop has a positive value when 
averaged over the lattice \cite{Cucchieri,Karsch:1994xh}.


\section{Lattice results}\label{slattice}

The propagators of the gluon and of the Faddeev-Popov ghost have been numerically
evaluated using the methods described in Refs.\
\cite{Cucchieri:2006tf,Bloch:2003sk,Cucchieri:2003di,Cucchieri:2006za}.
Details can also be found in Table \ref{conf}.
Finite temperature has been introduced according to the standard procedure of reducing 
the extent of the time direction compared to the spatial ones \cite{Rothe:1997kp}. 
We consider $N_t$ lattice sites along the temporal
direction and $N_s$ sites along the spatial direction with $N_s \gg N_t$.
After taking the infinite-spatial-volume limit $V_s = N_s^3 \to \infty$ ,
the continuum limit is given by $N_t \to \infty$ and $a \to 0$,  keeping the
product $a N_t$ fixed.
This yields a temperature 
\be
T\,=\,\frac{1}{a N_t} \; .
\label{eq:T}
\ee
In the following, we first consider the symmetric three-dimensional case,
discussing finite-volume effects in detail. We then move on to the four-dimensional
case and discuss the effects of finite temperature.


\subsection{Finite-volume effects}\label{svolume}

\begin{figure}
\includegraphics[width=\linewidth]{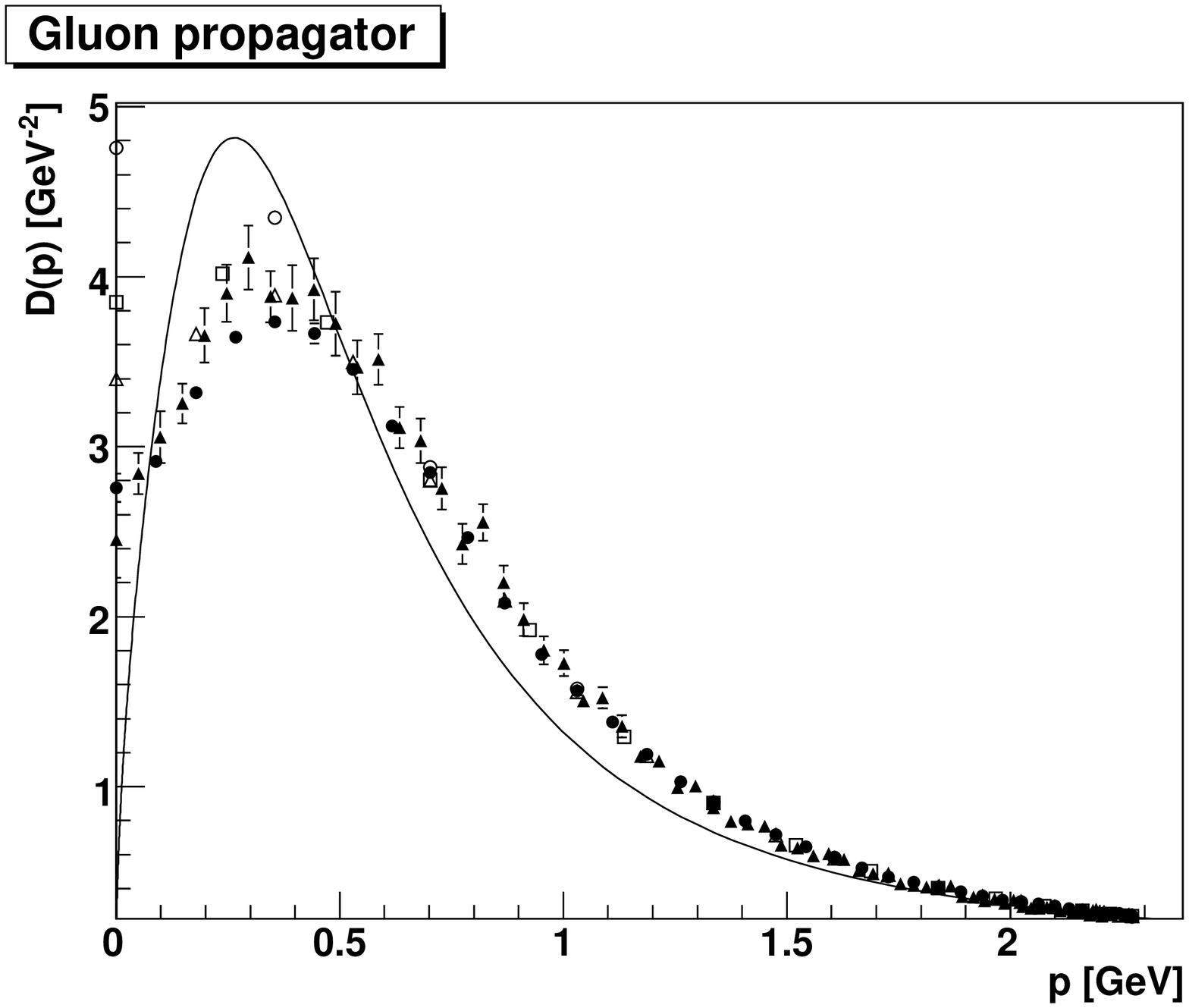}\\
\includegraphics[width=\linewidth]{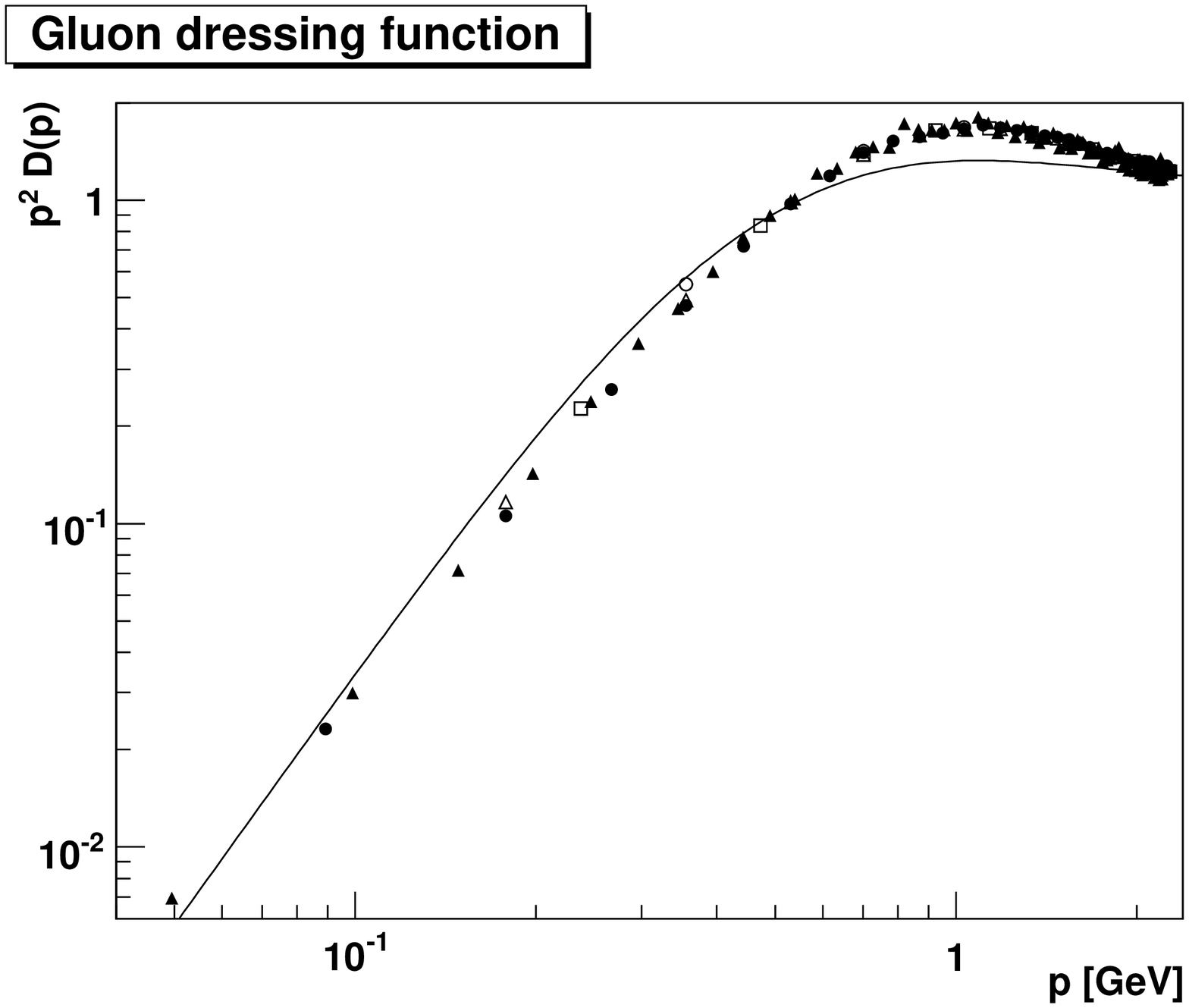}
\caption{\label{gpv}
The gluon propagator $D(p)$ (top) and dressing function $p^2 D(p)$ (bottom)
as a function of the momentum $p$ for various volumes in three dimensions.
Open circles, open squares, open triangles, full circles and full triangles
correspond to {\em Setups} 16 ($V= 20^3$), 17 ($V= 30^3$), 18 ($V= 40^3$), 19 ($V= 80^3$)
and 20 ($V= 140^3$), respectively. The solid line is the result from Dyson-Schwinger
calculations  \cite{Maas:2004se}, yielding a gluon propagator suppressed as $p^{0.59}$
in the IR limit.}
\end{figure}

\begin{figure}
\includegraphics[width=\linewidth]{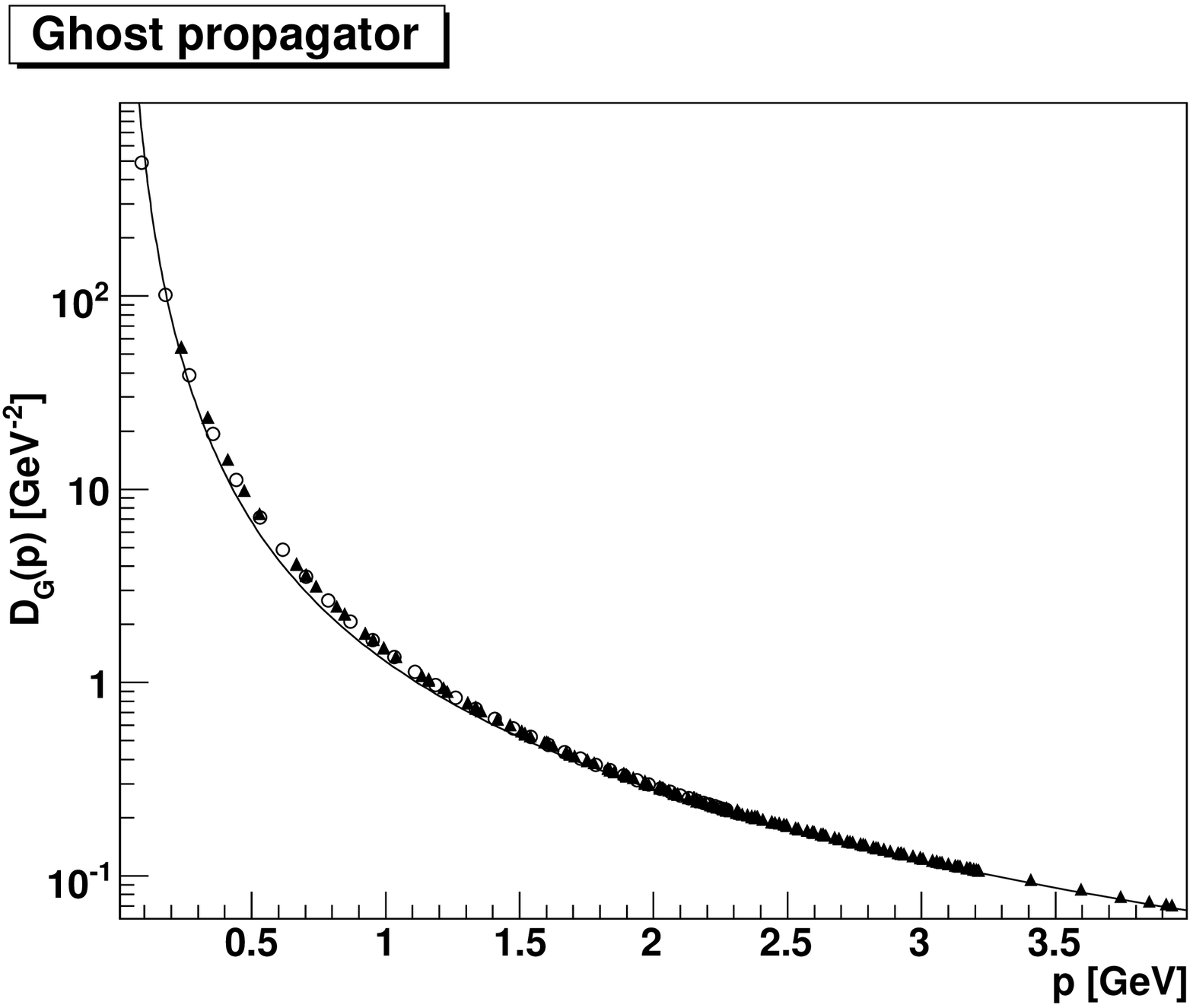}\\
\includegraphics[width=\linewidth]{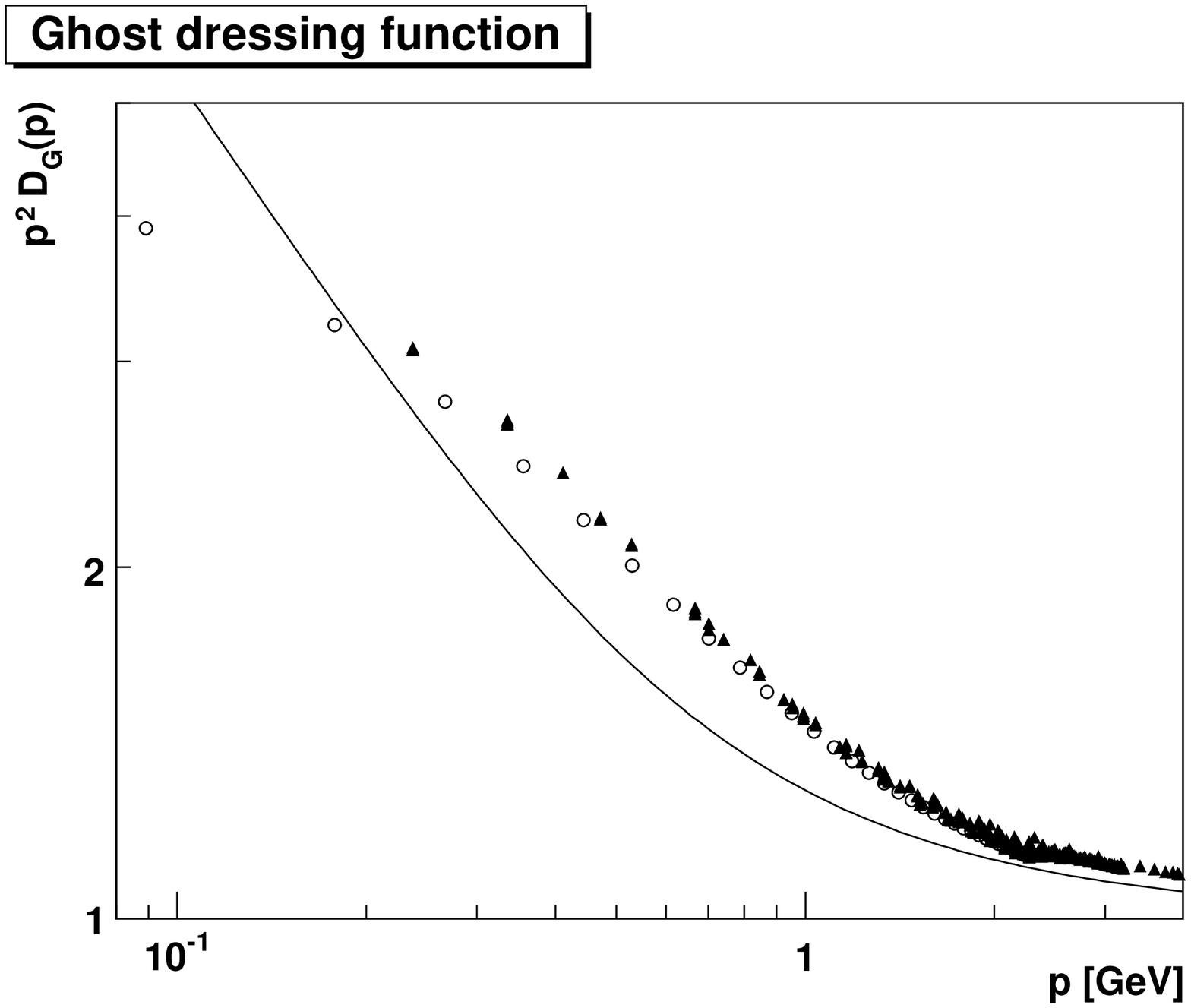}
\caption{\label{ghpv}
The ghost propagator $D_G(p)$ (top) and the dressing function $p^2 D_G(p)$
(bottom) as a function of the momentum $p$ for two different volumes in
three dimensions. Full triangles and open circles correspond to {\em Setups}
17 ($V = 30^3$) and 19 ($V = 80^3$), respectively. The solid line is the result
from Dyson-Schwinger calculations \cite{Maas:2004se}, yielding a ghost
propagator enhanced as $p^{-2.8}$ in the IR limit.}
\end{figure}

\begin{figure}
\includegraphics[width=\linewidth]{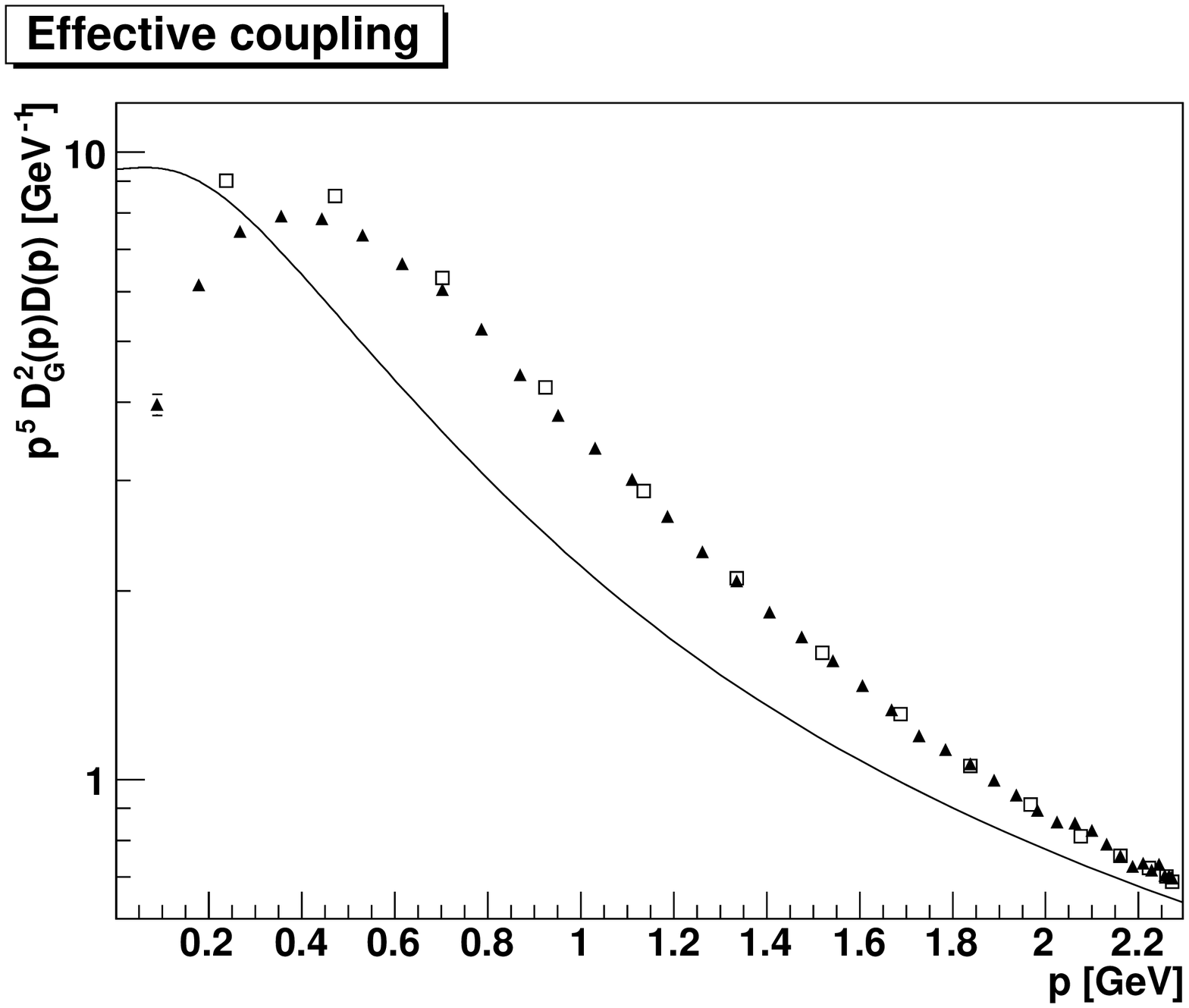}
\caption{\label{alpha}
The {\em effective coupling} $p^5 \, D_G^2(p)\, D(p)$ as a function of the momentum $p$
for {\em Setups} 17 (open squares, $V = 30^3$) and 19 (full triangles, $V = 80^3$).
The solid line is the result from the same Dyson-Schwinger calculations \cite{Maas:2004se}
reported in Figs.\ \ref{gpv} and \ref{ghpv}.}
\end{figure}

Before considering the finite-temperature dependence of the propagators,
it is worthwhile to discuss the effects of finite volume on their IR behavior.
Clearly, a finite volume $V$ affects all processes with a correlation length
$\xi$ of the order of (or larger than) $V^{1/d}$. [In the finite-temperature
case one can expect finite-volume effects when $\xi \gtrsim V_s^{1/(d-1)}$,
where $V_s$ is the spatial volume.] Since confinement is induced by an
infinite correlation length \cite{Alkofer:2000wg}, finite-volume effects are likely to
be observed in the numerical evaluation of gluon and ghost propagators.\footnote{One
should recall that finite-size effects are more or less pronounced depending on
the quantity considered. In particular, one does not expect such effects for
quantities with an intrinsic mass-scale $m$, such as quarks or hadrons,
since in these cases the relevant correlation length is of the order of $1/m$.
This has an important consequence in studies using functional methods: the
specific behavior in the deep IR region of the gluons is nearly irrelevant for
hadronic observables \cite{Alkofer:2000wg,Fischer:2006ub,Holl:2006ni} and results
are sensitive only to the behavior at intermediate momenta (about $0.5$--1 GeV).}
These effects should be considered when comparing the numerical data
to the predictions of the Gribov-Zwanziger and of the Kugo-Ojima confinement
scenarios.  Let us recall that several calculations in the continuum, e.g.\ using
DSEs \cite{Zwanziger,Alkofer:2000wg,Fischer:2006ub,vonSmekal} or the renormalization
group \cite{Gies}, are in favor of these scenarios and confirm their predictions.
In addition, using DSEs one can also study (in the continuum) the consequences of
a finite volume on these propagators. In particular, it has been found \cite{fischer}
that in a finite volume the gluon propagator is suppressed at small momenta
and finite at zero momentum, while the ghost propagator is IR enhanced.
These results are in qualitative agreement with the findings using lattice calculations
\cite{Bloch:2003sk,Ilgenfritz:2006gp,Ilgenfritz:2006he,Silva}.
Moreover, using DSEs it can be argued \cite{fischer} that an IR-finite
gluon propagator is indeed an artifact of the finite volume and that 
one obtains a null propagator at $p=0$ when the infinite-volume limit is considered.
On the lattice the situation is clearly more complicated. Even in the case of
a power-dependence of the type $D(0) \propto 1/V^b$, it can be very difficult
\cite{Cucchieri:2003di} to perform simulations with sufficiently large values of the 
physical lattice volume $V$ so
that one can really control the extrapolation to infinite volume. This is in fact the
case and up to now there is not yet any convincing result on a (symmetric)
four-dimensional lattice showing an IR-vanishing gluon propagator $D(p)$.
The situation is actually even worse than this. Indeed, in order to see an
IR-suppressed propagator one should obtain that $D(p)$ has a maximum for some
value of $p \neq 0$. Up to now this has not been obtained in the Landau 4d case
(using symmetric lattices), even when considering very large lattices
(see Fig.\ 10 in \cite{Cucchieri:2006xi} and Fig.\ 3 in \cite{Ilgenfritz:2006he}).
There are, however, evidences of a suppressed gluon propagator when considering
4d asymmetric lattices \cite{Silva}, even though in this case it is difficult to extract
quantitative information from the data \cite{Cucchieri:2006za,Ilgenfritz:2006gp}.
Also, a suppressed gluon propagator at small momenta can be obtained
in the 4d case when simulations are done in the strong-coupling regime
\cite{Cucchieri:1997fy}. Finally, a suppressed gluon propagator 
has recently been obtained 
(using symmetric lattices and $\beta$ values in the scaling region)
for a Landau-like gauge condition \cite{Cucchieri:2007uj}, confirming that the
difficulties in finding a similar IR-suppressed gluon propagator in Landau gauge
are probably related to finite-size effects.

\begin{figure*}
\includegraphics[width=\linewidth]{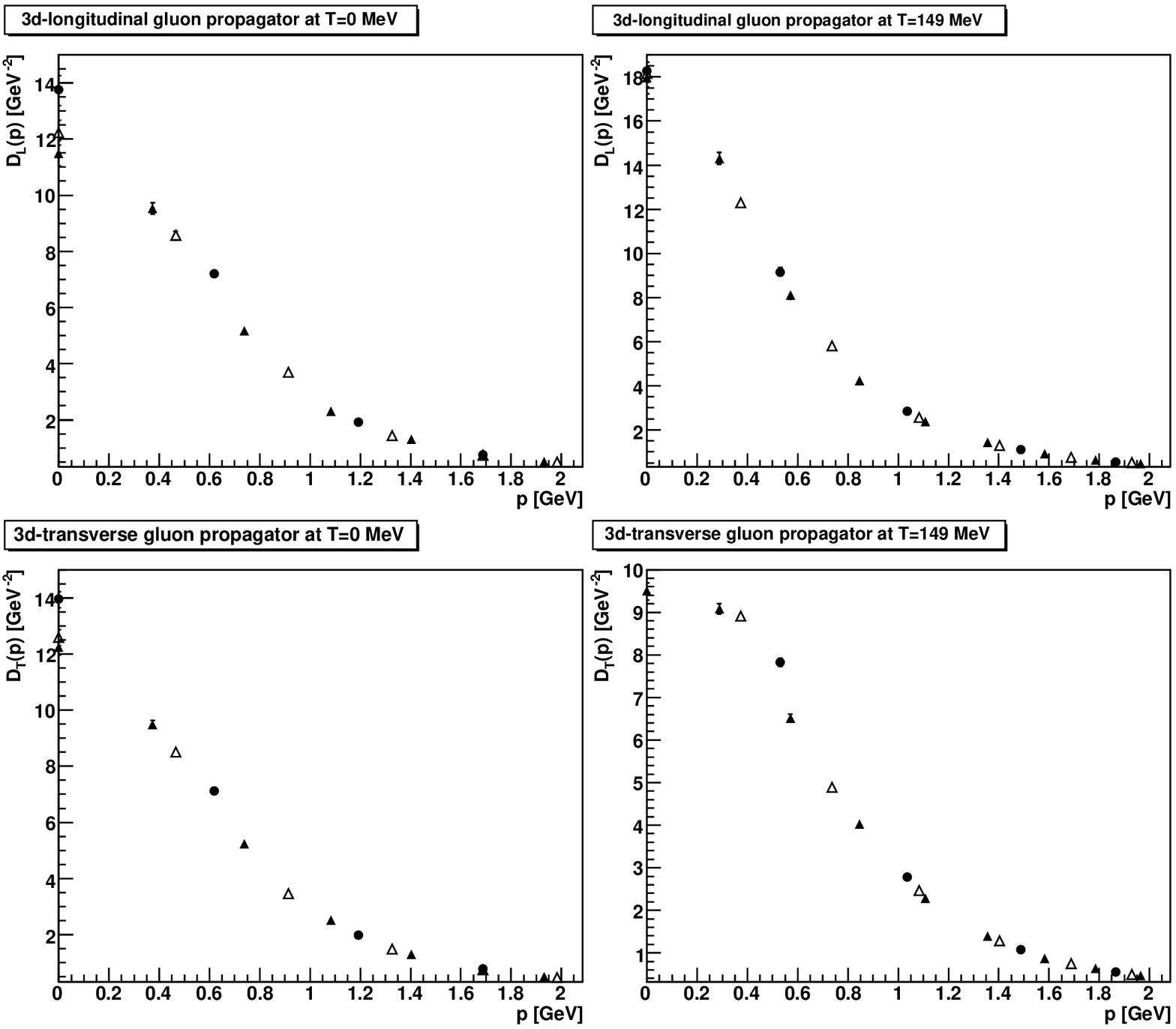}
\caption{\label{gp-vlow}
The volume dependence of the gluon propagators in the low-temperature phase.
The top panels show the 3d-longitudinal propagator while in the bottom panels
we report results for the 3d-transverse propagator. The left panels show
the results at zero temperature ($N_s = N_t$) and the right panels at $T\approx119$ MeV
($N_t = 10$),
well inside the low-temperature phase. We always compare results obtained
using three different physical volumes.
The smallest volumes ({\em Setups} 1 and 5) are indicated with full
circles, the middle-size volumes ({\em Setups} 2 and 6) are represented by
open triangles and the largest volumes ({\em Setups} 3 and 7) are
indicated by full triangles. Momenta are aligned along the $x$ axis.}
\end{figure*}

Numerically, it is clearly easier to consider first the three-dimensional case,
since one can then use much larger lattice sides.
Moreover, the Gribov-Zwanziger scenario can be applied
to three dimensions \cite{Zwanziger:1990by} --- as also supported by 
numerical studies \cite{Cucchieri,Cucchieri:1999sz,Cucchieri:2006tf,
Cucchieri:2003di} and by calculations using DSEs \cite{Zwanziger,Maas:2004se} ---
and the IR suppression of the gluon propagator is expected to be stronger than
in the 4d case (i.e.\ the IR exponent is larger in the 3d case).
Finally, since the theory is finite, renormalization issues do not obscure the
interpretation of the results. Let us recall that various lattice calculations
have been performed in the 3d case, both for the propagators
\cite{Cucchieri,Cucchieri:1999sz,Cucchieri:2006tf,Cucchieri:2003di}
and for the 3-point vertices \cite{Cucchieri:2006tf}. These results
support the assumptions usually made in DSE calculations \cite{Maas:2004se}.

We now consider symmetric lattices in the 3d case (see run parameters at the
bottom of Table \ref{conf}).
These can be interpreted as lattices at $T=0$ in 3d or as lattices
at $T=\infty$, simulating the dimensionally-reduced theory (without the
Higgs field). We compare the numerical data to
Dyson-Schwinger calculations \cite{Maas:2004se}.
In Fig.\  \ref{gpv} we present results in the 3d case for the gluon propagator using lattice
simulations and DSE calculations. Considering the lattice data on the larger
lattices, there is a clear maximum at $p_{max} \approx 400$ MeV. Moreover, as
the lattice volume increases, the IR suppression becomes stronger.
However, even using very large volumes (i.e.\ $V = 140^3$ \cite{Cucchieri:2003di}),
only few momenta are
available in the interval $[0, p_{max}]$. The comparison to the DSE solution
(for infinite volume) shows qualitative agreement.
Concerning the gluon dressing function $p^2 D(p)$, it is found that lattice
results exhibit a power-law behavior for momenta below $400-500$ MeV.
The result is very similar to the DSE solution, albeit differing by a constant factor.
Of course, when considering the dressing function $p^2 D(p)$,
the factor $p^2$ is dominant in the IR limit and the agreement between lattice
and DSE results looks better.
Also note that the pre-factor of the power-law
obtained in DSE studies is sensitive to the approximations performed \cite{Maas:2004se}.

\begin{figure*}
\includegraphics[width=\linewidth]{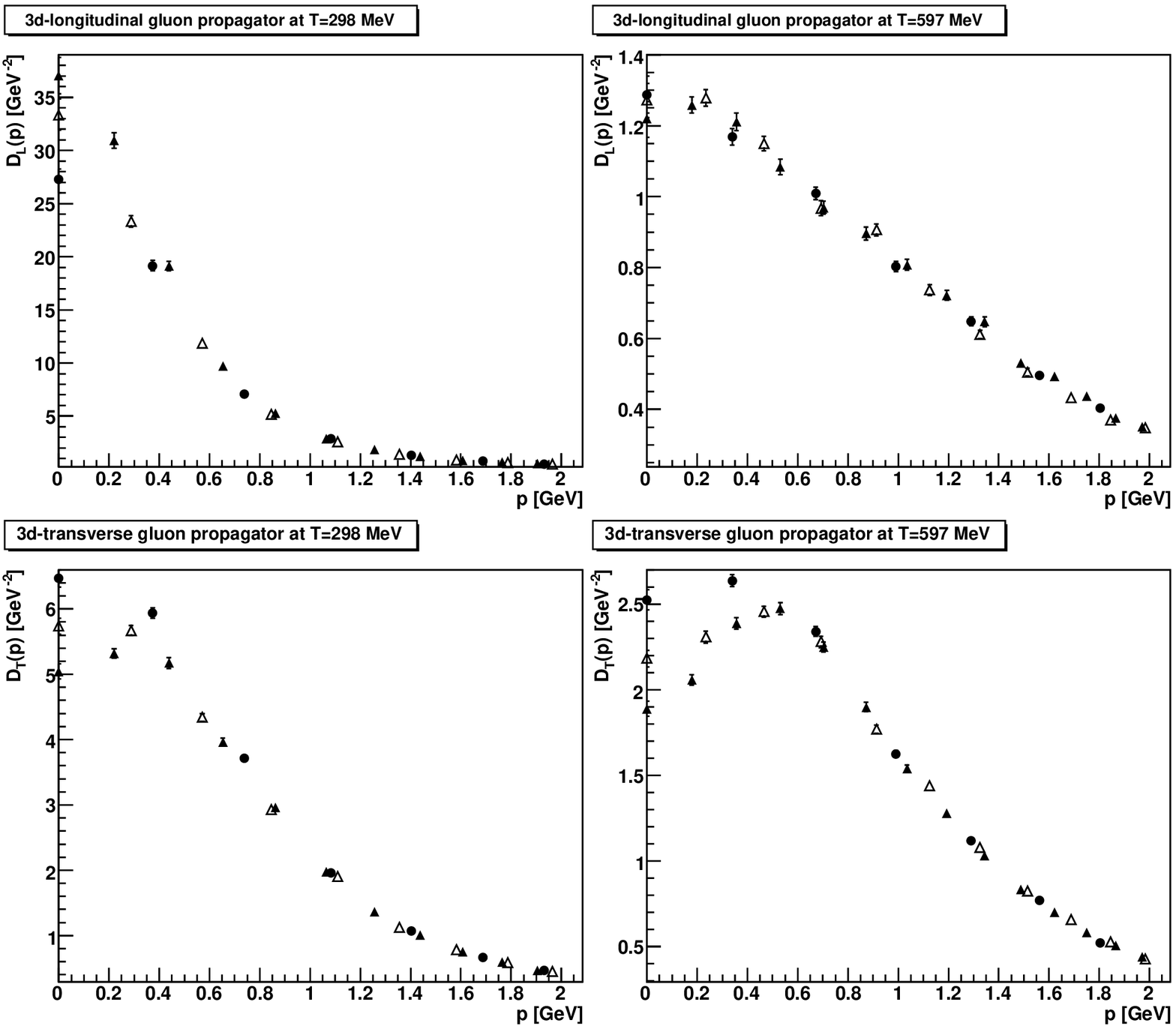}
\caption{\label{gp-vhigh}
The volume dependence of the gluon propagators in the high-temperature phase.
The top panels show the 3d-longitudinal propagator while in the bottom panels
we report results for the 3d-transverse propagator. The left panels show
the results at $T\approx298$ MeV ($N_t = 4$), slightly above the thermodynamic transition,
and the right panels at $T\approx597$ MeV ($N_t = 2$),
well inside the high-temperature phase. We always compare results obtained
using three different physical volumes.
The smallest volumes ({\em Setups} 8 and 13) are indicated with full
circles, the middle-size volumes ({\em Setups} 9 and 14) are represented by
open triangles and the largest volumes ({\em Setups} 10 and 15) are
indicated by full triangles. Momenta are aligned along the $x$ axis.}
\end{figure*}

Results for the ghost propagator $D_G(p)$ and the dressing function $p^2 D_G(p)$
are reported in Fig.\  \ref{ghpv}. In this case there are no evident finite-volume
effects in the lattice data for $D_G(p)$, even though 
such effects are visible at small momenta
when considering the dressing function $p^2 D_G(p)$.
In particular, the dressing function shows a reduced IR divergence for
the largest lattice at small momenta $p$. On the other hand, one should recall
that, for a relatively small statistics, the so-called ``exceptional configurations''
are probably not adequately sampled
and that they contribute significantly to the IR enhancement of $D_G(p)$, i.e.\
this IR enhancement could be underestimated
\cite{Sternbeck:2005tk,Cucchieri:2006tf}.
When considering the comparison to DSE results, we see that agreement
is again at the qualitative level
(see in particular the dressing function in the bottom figure). Indeed, the lattice
data show a weaker divergence than the one obtained using DSE calculations. As
said above, this could be related to an insufficient statistics for the ghost
propagator when large lattice volumes are used.

\begin{figure*}
\includegraphics[width=\linewidth]{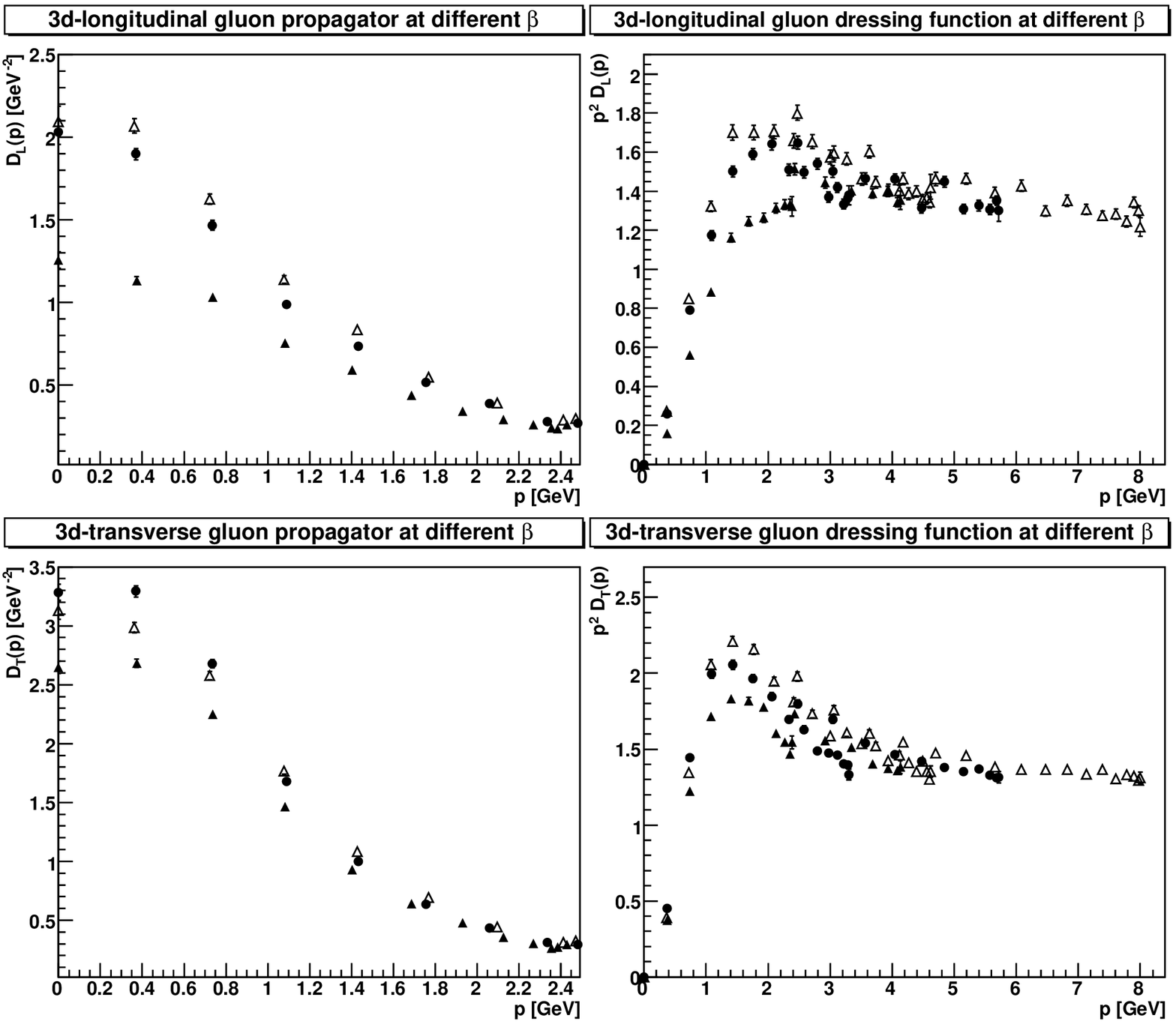}
\caption{\label{gp-beta}
The $\beta$ dependence of the gluon propagators for a physical spatial
volume $V_s$ of approximately $(3.35$ fm$)^3$ and temperature $T \approx 575 MeV$,
using three different lattice setups. We report data for the
3d-longitudinal propagator in the top-left panel, for the
3d-longitudinal dressing function in the top-right panel,
for the 3d-transverse propagator in the bottom-left panel and
for the 3d-transverse dressing function in the bottom-right panel.
We represent with full triangles data at $\beta=2.3$ ({\em Setup} 13),
with full circles data obtained at $\beta=2.4$ ({\em Setup} 11) and with
open triangles results for $\beta=2.5$ ({\em Setup} 12).
Momenta below approximately 2.4 GeV are considered along the $x$-axis, larger momenta are
along the 3d-spatial diagonal.}
\end{figure*}

Finally, one can consider the {\em effective coupling}\footnote{Since the 3d theory is
(ultraviolet) finite this quantity is of course not a running coupling in the
sense of the renormalization group as in the 4d case \cite{Alkofer:2000wg,
Fischer:2006ub}. Nevertheless, using DSE calculations, one can make a prediction
for its IR behavior also in the 3d case.}
defined to be proportional to $p^5 \,D_G^2(p)\, D(p)$, for which DSEs predict a constant limit at zero momentum.
The comparison of lattice data and DSE results is made in Fig.\  \ref{alpha}.
It is clear that the asymptotic (constant) regime is reached in the Dyson-Schwinger
result only for momenta of the order of 200 MeV. For the lattice data, when $V = 30^3$,
we do not have any point in the range $[0, 200]$ MeV and it is difficult to
say what one could find in the $p=0$ limit. On the other hand, in the case of the
larger lattice ($V = 80^3$), one sees a coupling decreasing when $p$ is smaller
than about $p_{max} \approx 400$ MeV. 
It thus seems that the two approaches differ qualitatively in the deep IR limit.
Nevertheless, it is interesting to note that the maximum of the data for the larger
lattice is obtained for the
momentum $p = p_{max} \approx 400$ MeV,  where also the gluon propagator reaches its
maximum value. Thus, a possible interpretation \cite{Cucchieri:2006xi,fischer}
is that finite-size effects are different for the two propagators and one
might see an IR-finite effective coupling only in the infinite-volume limit.
In other words, since the prediction from DSE studies is based on the
relation\footnote{Here we define the IR exponents $a_D$ and $a_G$
using the relations $D(p) \sim p^{2 a_D - 2}$ and $D_G(p) \sim p^{-2 a_G -2}$.}
$a_D - 2 a_G = (4 - d)/2$ \cite{Zwanziger,Lerche:2002ep}, one can
imagine that the true values for the exponents $a_D$ and $a_G$ are obtained only in
the infinite-volume limit and that at finite volume the above relation
does not need to be satisfied.\footnote{For a different interpretation, see \cite{Boucaud:2006if}.}

Let us note that Dyson-Schwinger studies find that the expected power-law behaviors
for the propagators (in 3d as well as in 4d) start to appear only for momenta
smaller than some energy-scale $\Lambda_I\lesssim 200$ MeV \cite{fischer}.
Also, one should consider with caution lattice results at the smallest non-zero 
momentum $\,2/a\sin(\pi/N) \approx 2\pi/(aN)$, i.e.\ when the quantities considered
may ``feel'' the boundaries. Thus, one should try to extract the IR behavior of the
propagators only using data in the range
\be
\frac{2}{a}\sin\left(\frac{\pi}{N}\right)\ll p\lesssim \Lambda_I
\label{valrange}.
\ee
With a lattice spacing $a$ of order of 1 GeV$^{-1}$ one consequently needs $N \gg 50$.
Present simulations have access to this region only in the three-dimensional case.
In the four-dimensional case the situation is more complicated and,
in particular, since the IR suppression of the 4d gluon propagator is predicted
to be weaker than in three dimensions ($p^{0.38}$ instead of $p^{0.59}$) \cite{Zwanziger,
Lerche:2002ep}, one probably needs even larger lattice sides than in the 3d case,
making these computational studies very demanding. Looking at the results obtained
in the 3d case and considering the data obtained with the largest 4d
lattices available \cite{Cucchieri:2006xi,Ilgenfritz:2006he}, one can argue that
volumes larger by a factor of 10 (i.e.\ a factor of about 2 in lattice side) are
probably necessary in order to see a decreasing gluon propagator also in the
4d case. On the other hand, for the ghost propagator and with the available
lattice sides, the IR-enhancement is clearly observed \cite{Bloch:2003sk,
Ilgenfritz:2006he,Silva,Ilgenfritz:2006gp}.
even though the value obtained numerically for the IR exponent $a_G$ is usually
smaller than the DSE result \cite{Zwanziger,Lerche:2002ep}.


\subsection{Finite temperature}

\begin{figure*}
\includegraphics[width=\linewidth]{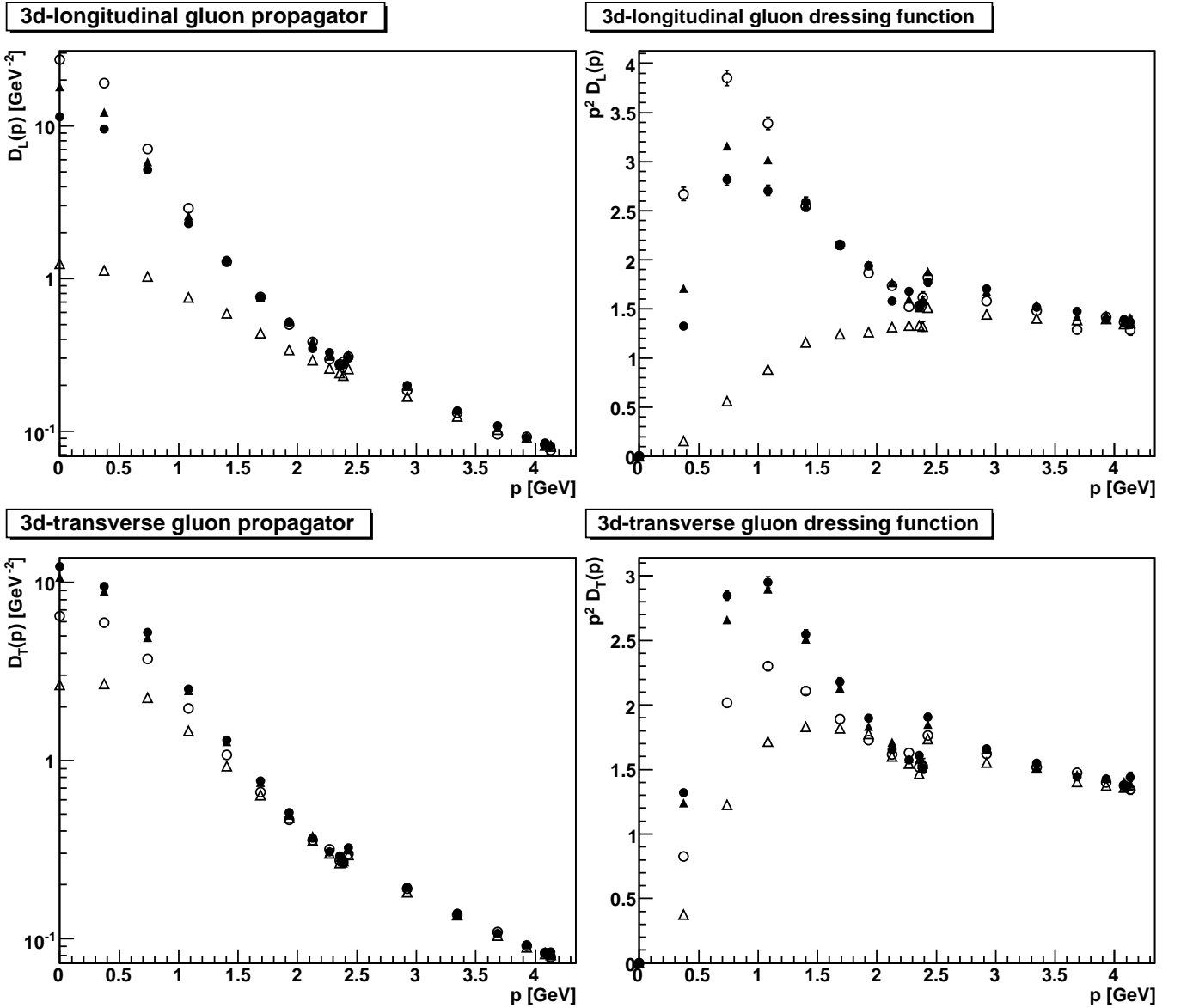}
\caption{\label{fgp-sv}
The 3d-longitudinal and 3d-transverse
gluon propagators and dressing functions as a function of the
temperature for the same spatial volume $V_s$ of approximately $(3.3$ fm$)^3$.
We report data for the
3d-longitudinal propagator in the top-left panel and for the
3d-longitudinal dressing function in the top-right panel.
Results for the 3d-transverse propagator are shown in the bottom-left panel and
for the 3d-transverse dressing function in the bottom-right panel.
Full circles, full triangles, open circles, and open triangles correspond
to {\em Setups} 3 ($T=0$ MeV), 6 ($T\approx119$ MeV), 8 ($T\approx298$ MeV), and 13
($T\approx597$ MeV), respectively.  Momenta respectively below and above 2.4 GeV are measured 
along the $x$-axis and along the spatial diagonal. 
The break at 2.4 GeV is due to violation of rotational invariance.}
\end{figure*}

\begin{figure*}
\includegraphics[width=\linewidth]{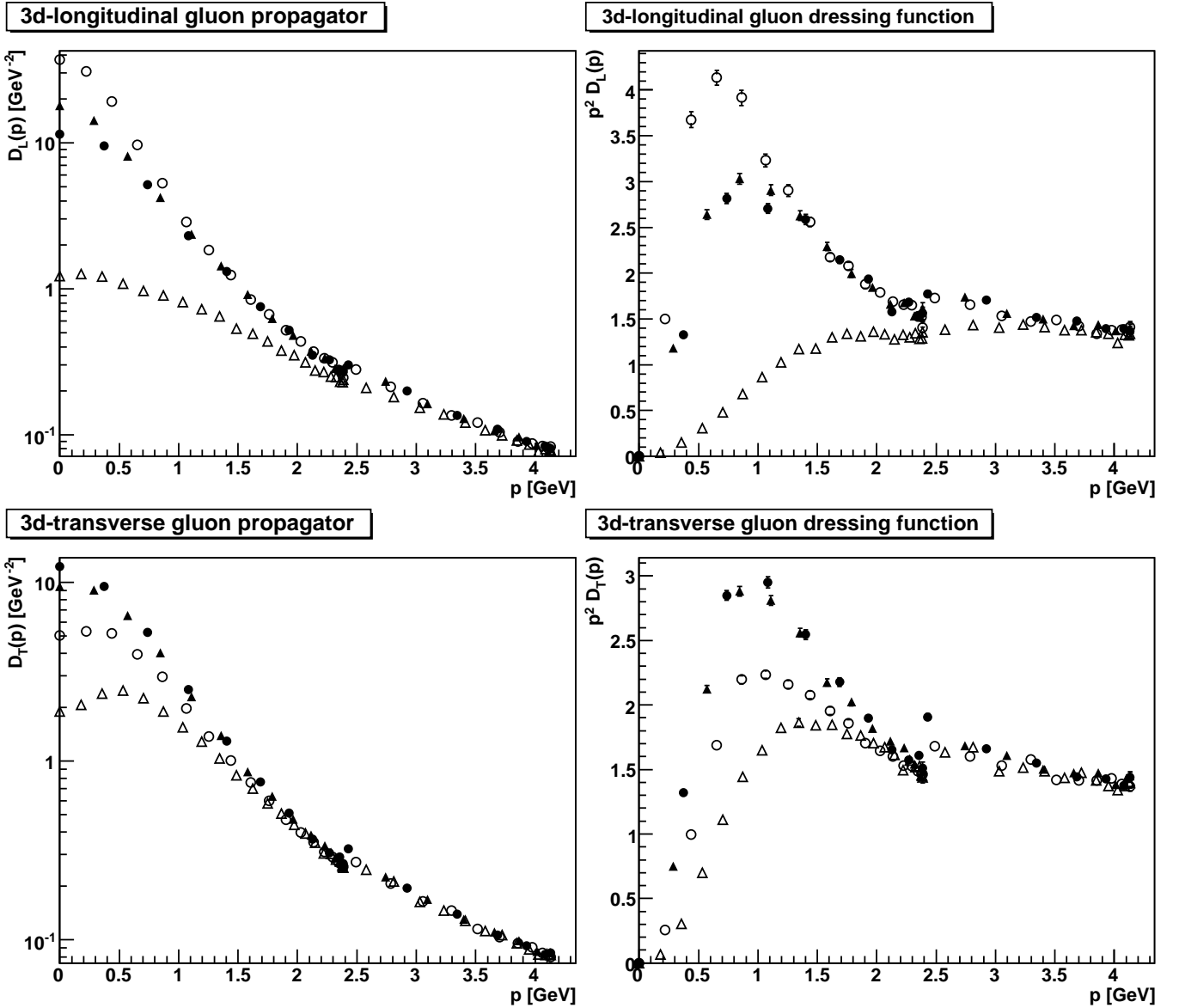}
\caption{\label{fgp}
As figure \ref{fgp-sv}, but considering for each temperature $T$ the largest 
spatial volume available.
Full circles, full triangles, open circles, and open triangles correspond
to {\em Setups} 3 ($T=0$ MeV), 7 ($T\approx119$ MeV), 10 ($T\approx298$ MeV) and 15
($T\approx597$ MeV), respectively.}
\end{figure*}

\begin{figure}
\includegraphics[width=\linewidth]{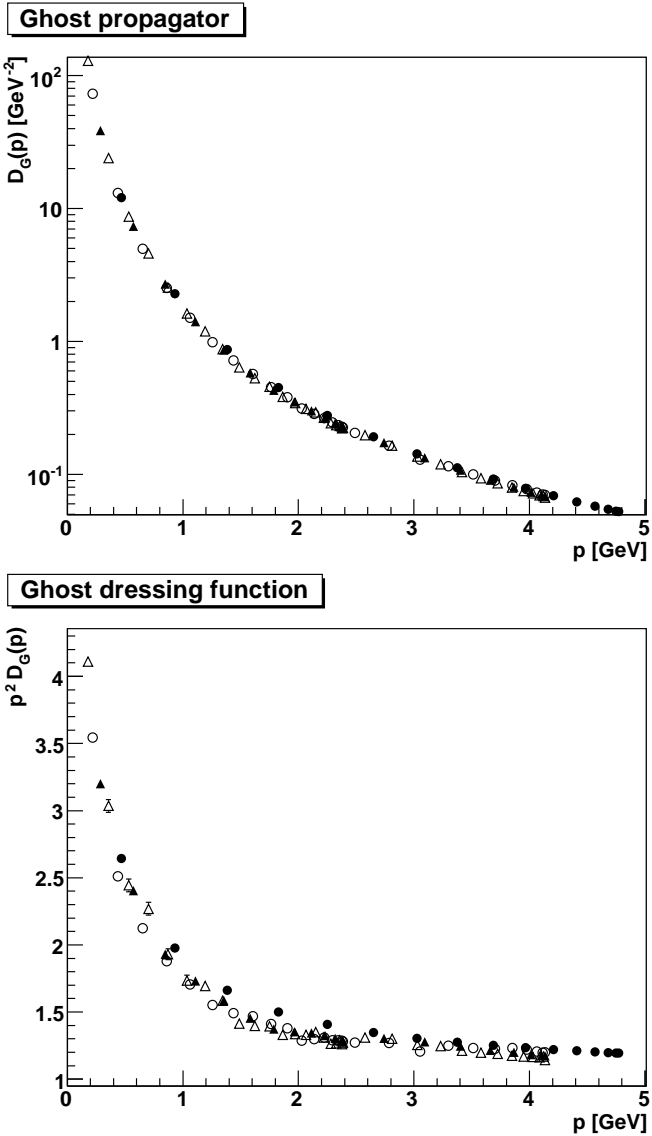}
\caption{\label{fghp}
The ghost propagator and dressing function as a function of temperature.
The top panel shows the ghost propagator and the bottom panel the
corresponding dressing function. 
Full circles, full triangles, open circles, and open triangles correspond
to {\em Setups} 4 ($T=0$ MeV), 7 ($T\approx119$ MeV), 10 ($T\approx298$ MeV) and 15
($T\approx597$ MeV), respectively.}
\end{figure}

\begin{figure}
\includegraphics[width=\linewidth]{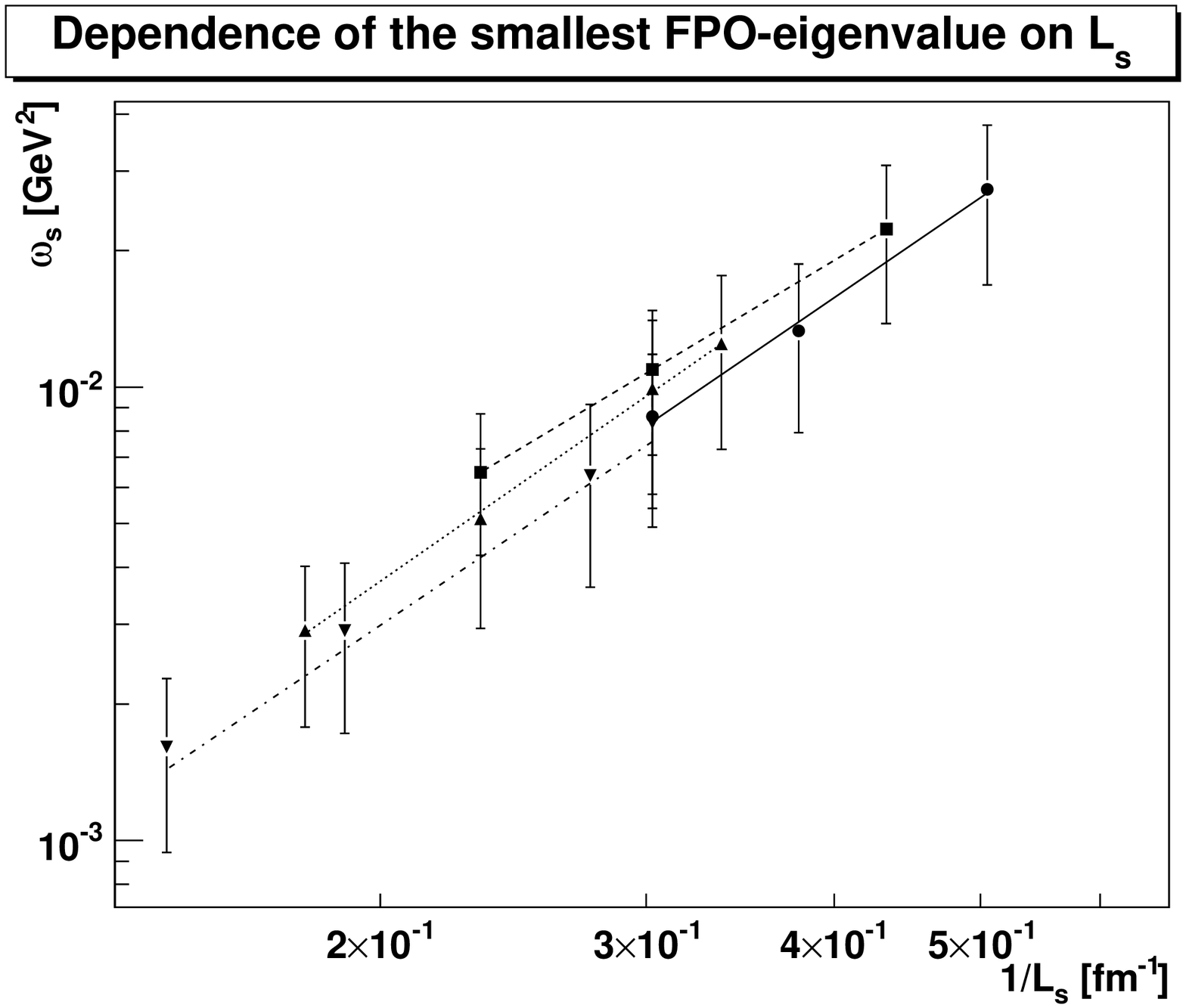}\\
\includegraphics[width=\linewidth]{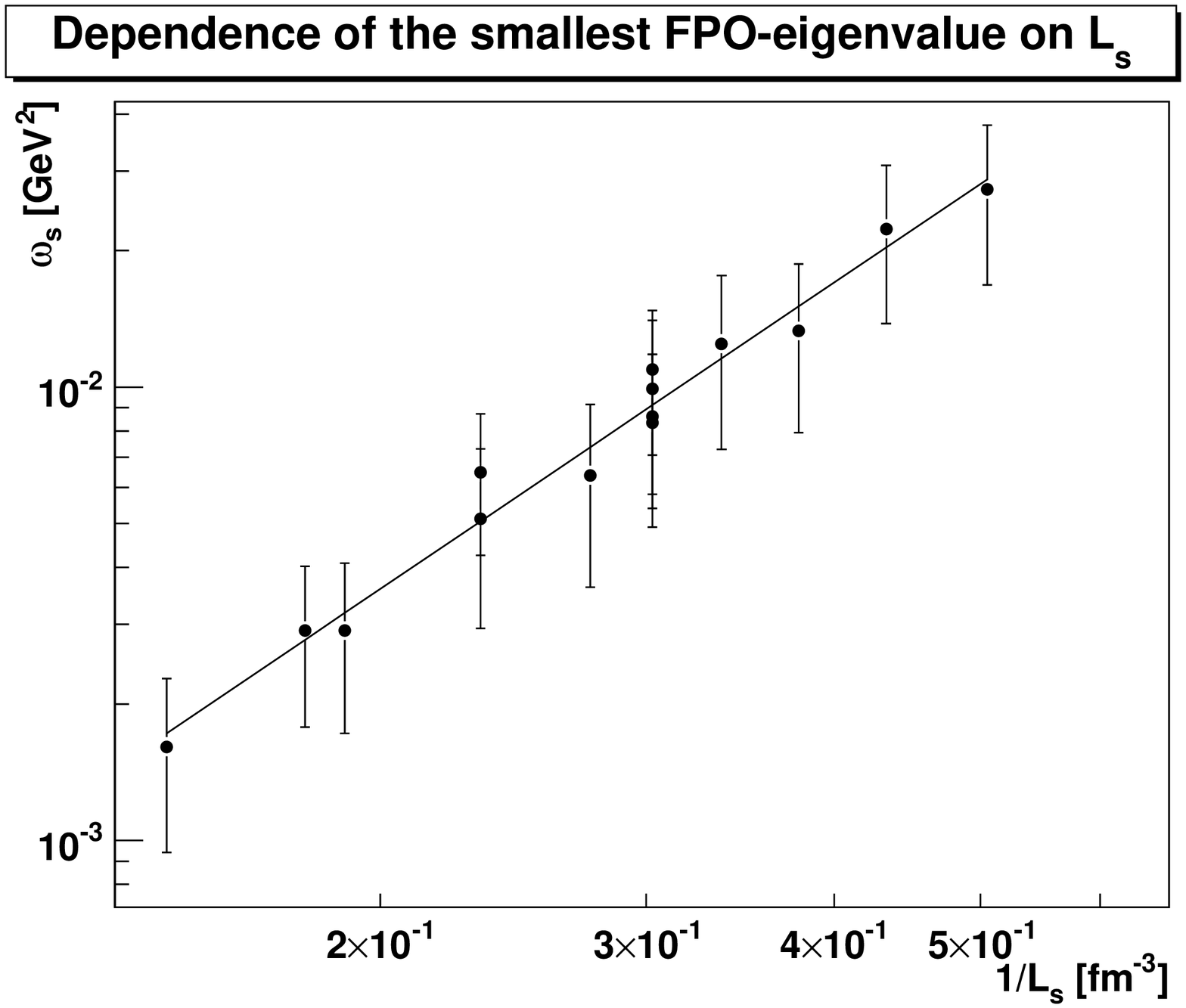}
\caption{\label{seigen}
The dependence of the lowest eigenvalue $\omega_s$ of the FPO on the spatial 
extension $L_s$. The lines are fits of the type
$cx^{-b}$. In the top figure, we show fits for each temperature separately.
The fit parameters can be found in Table \ref{fittable}.
Solid lines and circles are at $T=0$ MeV, dashed lines and squares are at
$T\approx 119$ MeV, dotted lines and triangles are at $T\approx 298$ MeV while dashed-dotted
lines and upside-down triangles are at $T\approx 597$ MeV. In addition to the {\em Setups}
listed in Table \ref{conf} we obtained some of the data using lattice volumes 4$\times$18$^3$ and
$2\times 22^3$ at $\beta=2.3$.
In the bottom figure we consider data for all the temperatures.}
\end{figure}

Here we present our calculations for gluon and ghost propagators in 4d at 
finite temperature. 
We consider four different temperatures, two of them below the thermodynamic 
transition (i.e.\ at zero temperature and at $T \approx 119$ MeV) and two above it
(i.e.\ one at $T \approx 298$ MeV and the other at $T \approx 597$ MeV).
Our runs are summarized in Table \ref{conf}.
Note that, for a time extension $N_t = 4$, the critical coupling is
$\beta=2.299$ \cite{Fingberg:1992ju}, corresponding to $T_c\approx 295$ MeV. 
Here the calculations for this time extension have been done at $\beta=2.3$.
Thus, the temperature
$T \approx 298$ MeV is just above the thermodynamic transition, while
the highest temperature ($T \approx 597$ MeV) corresponds to about
twice the critical temperature $T_c$.
This allows us to make contact with previous studies \cite{Cucchieri}
of the high-temperature phase, which investigated the gluon propagator
for $T \gtrsim 2\, T_c$.
In order to clarify possible finite-volume effects, we considered for
each temperature three different spatial volumes. Due to
memory limitations, the spatial sizes are chosen to depend on the
temperature, the largest ones being at the highest temperatures,
for which $N_t$ is smallest.
We find that, in the ghost case, neither the propagator nor the dressing function
show any visible volume or discretization effects within the statistical errors.

The volume dependence of the gluon propagators for the two {\em Setups} below
the thermodynamic transition is shown in Fig.\  \ref{gp-vlow}.
As expected, at zero temperature, $D_T(p)$ and $D_L(p)$ coincide.
In this case no clear volume dependence is observable,
since the three spatial volumes are rather similar.
At $T \approx 119$ MeV, $D_T(p)$ and $D_L(p)$ are already quite different
(see discussion below for details). On the other hand,
these two functions still do not show any strong volume dependence.
Also note that the consequences of the violation of rotational symmetry are not
stronger than in the zero-temperature case \cite{Cucchieri:2006tf,Bloch:2003sk}.

The situation is different when considering the high-temperature phase
(see Fig.\  \ref{gp-vhigh}).
Indeed, at $T \approx 298$ MeV,
the two propagators show a clear, even though different,
volume dependence. In particular, the 3d-longitudinal gluon propagator
is enhanced at small momenta with increasing volume, while the
3d-transverse propagator becomes weaker in the IR with increasing volume.
Finally, at $T \approx 597$ MeV, the volume dependence of
the two propagators is similar, i.e.\ they both become smaller (in the IR region)
as the volume increases. Moreover, the 3d-transverse propagator
shows a distinct maximum at $p \approx 500$ MeV for the two largest volumes,
while the 3d-longitudinal propagator seems to go to a constant
at small $p$.
These IR behaviors at $T \approx 2 T_c$ are in agreement with
previous studies \cite{Cucchieri}.
Note that the apparently stronger finite-size effects seen at $T \gtrsim T_c$
compared to the low-temperature cases are likely due to the
larger spatial volumes (computationally) accessible in the
high-temperature phase. Also note that a maximum in the
3d-transverse propagator in the high-temperature phase is
observed already for a (spatial) lattice volume of $32^3$. At 
$T=0$, in the 4d case, one sees a clear maximum for $V \lesssim 32^4$
only when simulations are done in the strong-coupling regime
\cite{Cucchieri:1997fy}.
Thus, finite-size effects for the transverse propagator
seem to be smaller in the high-temperature phase, in agreement
with previous results.

In Fig.\  \ref{gp-beta} we check for discretization effects
by comparing results for three different $\beta$ values for
{\em Setups} that have, within a few percent, the same temperature
(i.e.\ $T \approx 575$ MeV) and the same physical spatial volume 
$V_s \approx (3.35$ fm$)^3$. 
The two gluon propagators show a different behavior.
Indeed, the 3d-transverse propagator is only weakly
affected by the $\beta$ value at the smallest momenta, the
effect being of the order of about 20 percent at most.
On the other hand, for the 3d-longitudinal propagator,
the effects in the IR are more pronounced, with variations
of up to a factor of almost 2.
This indicates that the scaling limit has not been reached yet
for these {\em Setups} in the IR region. Thus,
results for the 3d-longitudinal gluon propagator
at momenta below 1.5 GeV should be taken with caution.
Considering also the finite-volume effects discussed above,
one probably needs to consider larger and finer lattices
especially in the case of the 3d-longitudinal gluon propagator. 

We can now compare
the gluon propagators at different temperatures $T$.
In particular, in Fig.\  \ref{fgp-sv} we consider {\em Setups} with the
same physical spatial volume, while in Fig.\  \ref{fgp} we show data
using always the largest spatial volume available for a certain temperature.
The results are similar in the two cases and in the
the following we will mainly refer to Fig.\  \ref{fgp}.
One clearly sees that at sufficiently large momenta the
propagators at $T > 0$ coincide with the ones at zero temperature.
This is expected, since for $p \gg T$ zero-temperature perturbation
theory should re-emerge.
At smaller momenta, the 3d-transverse gluon
propagator decreases monotonically as the temperature
increases. Also, there is no sign of sensitivity to the
thermodynamic transition. 
The situation is different for the 3d-longitudinal propagator.
Indeed, in the IR region, it increases with $T$ when $T \lesssim T_c$,
while it strongly decreases when going from $T \approx T_c$ to
$T \approx 2 \, T_c$. From these results it is tempting to assume
that the 3d-longitudinal gluon propagator is largest
at the thermodynamic transition. However, due to the strong
finite-volume and discretization effects discussed above, a
confirmation of this conjecture requires more systematic studies
on larger and finer lattices.
Note that, due to renormalization effects, all dressing functions 
should vanish logarithmically for sufficiently large momenta.
Clearly this ultraviolet-asymptotic regime is not reached yet
with our range of momenta.

Finally, from Fig.\ \ref{fghp} we see that the Faddeev-Popov-ghost
propagator does not show any visible temperature dependence for all
temperatures $T$ considered here, in qualitative agreement with 
previous exploratory studies \cite{Langfeld:2002dd}.
We have also studied the spectrum of the Faddeev-Popov operator (FPO) ${\cal M}$ as a
function of the temperature $T$. In particular, we checked if and at what rate
the lowest eigenvalue $\omega_s$ of ${\cal M}$ vanishes in the infinite-volume
limit \cite{Zwanziger}. We recall that, in the 3d case, this eigenvalue
vanishes faster than the lowest eigenvalue of the Laplacian \cite{Cucchieri:2006tf}.
The behavior of $\omega_s$ as a function of the
spatial length $N_s a = L_s=(V_s)^{1/3}$ 
is shown in Fig.\ \ref{seigen}. We fitted our data using a power-law
function $c x^{-b}$. Results of these fits are reported in Table \ref{fittable}.
We see that the exponent $b$ varies little with $T$
and is greater than or equal to $2$, i.e.\ the eigenvalue goes to
zero at least as fast as the lowest eigenvalue of the Laplacian.
The magnitude $c$ is also essentially unaffected by the temperature.
This implies that, for
the lattice volumes and the lattice spacings considered here, the eigenvalue
$\omega_s$ (as a function of $1/L_s$) is already in the scaling region.
By fitting all data we get an exponent $b = 2.3(1)$
(see bottom of Fig.\ \ref{seigen}),
to be compared with the exponent of approximately 2.6 in the 3d case \cite{Cucchieri:2006tf}.

Summarizing our results for the propagators 
we find that 1) the ghost propagator does not have any
noticeable temperature dependence, 2) the 3d-transverse propagator depends on the
temperature, but it has the same (IR-suppressed) behavior below and above
the thermodynamic transition and 3) the 3d-longitudinal gluon
propagator does show a dependence on the temperature, with a different
behavior above and below the phase transition and an apparent IR divergence at $T_c$.
Moreover, at large $T$, the 3d-transverse propagator is suppressed
in the IR limit,
while the 3d-longitudinal one seems to behave
as a massive particle, approaching a plateau for small $p$. 
Indeed, for all temperatures, volumes and
discretizations, the 3d-longitudinal gluon dressing function vanishes in the IR.
This implies that the 3d-longitudinal propagator is always less
divergent than that of a massless particle. However, in order to check if the
behavior is indeed that of a {\em massive} particle, one needs much larger
physical volumes.
Let us stress that the results obtained here at the highest temperature
(i.e.\ $T \approx 2 T_c$)
agree qualitatively with those found in the dimensionally reduced theory
\cite{Cucchieri:1999sz,Cucchieri:2003di,Cucchieri:2006tf} and with previous studies of
gluon propagators at high temperature \cite{Cucchieri}.

A possible explanation of the above results will be discussed in Section \ref{sscenario}.

\begin{table}
\caption{\label{fittable}
Parameters for the fits in Fig.\ \ref{seigen}.
Fits have been done with {\tt gnuplot}.}
\begin{tabular}{|c|c|c|}
\hline
\hline
$T$ [MeV] & $c$ [GeV$^2$] & $b$ \cr
\hline
0 & 0.13(2) & 2.3(2) \cr
\hline
119 & 0.119(1) & 2.00(1) \cr
\hline
298 & 0.15(1) & 2.31(6) \cr
\hline
597 & 0.11(1) & 2.19(5) \cr
\hline
\hline
\end{tabular}
\end{table}


\section{Dyson-Schwinger equations}
\label{DSEq}

Due to the strong finite-volume effects observed in the numerical data, it is important
to complement the previous results with ones from the continuum and in the infinite volume,
e.g.\ using DSEs. Let us recall that studies at finite $T$ using
DSEs have already been presented in \cite{Maas:2005hs,Gruter:2004bb}. However, in both
cases, in addition to the truncations usually employed at zero temperature
\cite{vonSmekal}, various approximations have also been performed in
the solutions of the DSEs. Here, these additional approximations will
be removed. In this section we discuss the (analytic) asymptotic IR solutions of
the DSEs at any finite temperature, always assuming that external 
momenta $k$ obey $k\ll \Lambda_{\mathrm{QCD}}$. A (numerical) 
solution for all momenta, using
the same truncation scheme as in \cite{Maas:2005hs,Gruter:2004bb},
is presented in Appendix \ref{safull}. Let us stress that the
analysis presented here and in the Appendices \ref{sakernels}--\ref{safull} applies
to any semi-simple gauge group without any qualitative change.

The setup for the system of DSEs at finite $T$
has already been extensively discussed elsewhere \cite{Maas:2005hs,Gruter:2004bb,Maas:2005rf}.
Let us recall that, in the far infrared, one can use the so-called ghost-loop-only
truncation \cite{Maas:2005rf}, which keeps only the leading infrared term arising
in the derivation of the DSEs from the Faddeev-Popov-ghost term in the gauge-fixed Lagrangian. 
In the Gribov-Zwanziger scenario, this term dominates the action at low momenta.

\begin{figure}
\includegraphics[width=0.96\linewidth]{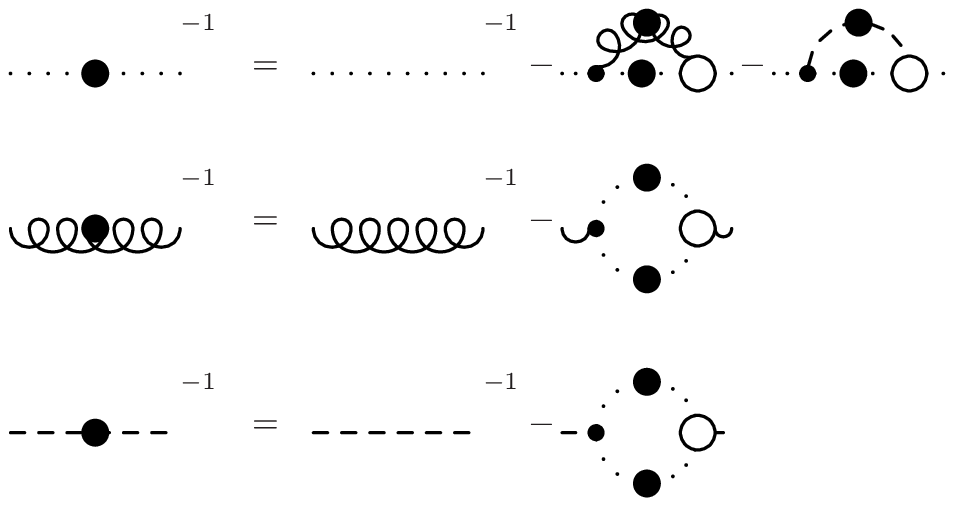}
\caption{\label{ffs}
The DSEs in the ghost-loop-only approximation at finite temperature.
Dotted lines represent ghosts, curly lines are used for the 3d-transverse gluons
and dashed lines represent 3d-longitudinal gluons.
Lines with a dot indicate a full propagator instead of a bare one.
Vertices with a small dot are bare ones, while those with a large, empty
dot represent full vertices. However, in our calculations
these vertices are also taken to be bare.}
\end{figure}

Then, the equations for the zero (soft, $k_0 = 0$) modes are given by \cite{Maas:2005rf}
\begin{widetext}
\bea
\!\!\!\!\!\!\!\!\! \frac{1}{G(0,\vec k)} &=& \tilde{Z}_3\, + \,
   \frac{g^2TC_A}{(2\pi)^3}\sum_{q_0}\int dq \, d\theta \,  \left[ A_T(0,q_0,\vec k,\vec q)
              \, G(q_0,\vec q) \, Z(q_0,\vec q-\vec k)
   \,+\,A_L(0,q_0,\vec k,\vec q)G(q_0,\vec q)H(q_0,\vec q-\vec k) \right]
\label{geq} \\
\!\!\!\!\!\!\!\!\! \frac{1}{Z(0,\vec k)} &=& Z_{3T}\, + \,
    \frac{g^2TC_A}{(2\pi)^3}\sum_{q_0}\int dq \, d\theta \, 
       R(0,q_0,\vec k,\vec q) \, G(q_0,\vec q) \, G(q_0,\vec q+\vec k)
\label{zeq}\\
\!\!\!\!\!\!\!\!\! \frac{1}{H(0,\vec k)} &=& Z_{3L}\, + \,
    \frac{g^2TC_A}{(2\pi)^3}\sum_{q_0}\int dq \, d\theta \,
       P(0,q_0,\vec k,\vec q) \, G(q_0,\vec q) \, G(q_0,\vec q+\vec k)
\label{heq} \; .
\eea
\end{widetext}
Here, we use the definitions $G(k^2) = k^2 D_G(k^2)$, $Z(k^2) = k^2 D_T(k^2)$,
$H(k^2) = k^2 D_L(k^2)$, where $D_G(k^2)$, $D_T(k^2)$ and $D_L(k^2)$ are, respectively,
the ghost propagator and the transverse and longitudinal gluon propagators.
The integral kernels $A_T$, $A_L$, $R$, $P$ and the angle $\theta$ are defined in Appendix
\ref{sakernels}. Also, $T$ is the temperature, $g$ is the bare coupling constant
and $C_A$ indicates the adjoint Casimir of the gauge group [for $SU(2)$, $C_A=2$].
The graphical representation of this system of equations is given in Fig.\ \ref{ffs}.
Note that Eqs.\ \prefr{zeq}{heq} have been obtained from the tensor equation for the gluon
propagator by contraction with generalizations of the projectors defined in Eqs.\
\pref{projt} and \pref{projl}. These tensors are parameterized by the
variables $\zeta$ and $\xi$ \cite{Maas:2005hs}, which appear explicitly only in the
integration kernels $R$ and $P$ (see Appendix \ref{sakernels}). Variations of these
parameters can be used to investigate truncation artifacts and, in particular, the
appearance of spurious divergences.\footnote{See e.g.\ \cite{Maas:2004se,Maas:2005hs,
Maas:2005rf} for a detailed discussion of this topic at finite temperature.}
Let us also note that the equations above reduce to the corresponding equations
of a 3-dimensional Yang-Mills-Higgs system when the hard modes ($q_0\neq 0$) inside loops
are neglected \cite{Maas:2004se}. Then, due to the ghost-loop-only approximation,
the Higgs field $H(0,\vec k)$ decouples\footnote{Indeed, it is easy to check that for
$k_0 = q_0 = 0$ one finds $A_L = P = 0$.} and $G(0,\vec k)$ and $Z(0,\vec k)$ become
the dressing function of a pure 3d Yang-Mills theory \cite{Maas:2004se}. In Appendix
\ref{avac} we show that the solutions of Eqs.\ \prefr{geq}{heq} at zero temperature
coincide with the well-known solutions presented in Refs.\ \cite{vonSmekal}.
Finally, we remark that one can find different wave-function renormalization constants
($Z_{3T}$ and $Z_{3L}$) in the gluon equations due to possible
finite contributions arising at non-zero temperature.

Equations \prefr{geq}{heq} can also be written as
\bea
\frac{1}{G(0,\vec k)} &=& \tilde{Z}_3 \,+\, \Pi_G(0,\vec k) \,+\,
                   \sum_{q_0\neq 0}\Pi_G(q_0,\vec k) 
\label{eq:geq1} \\
\frac{1}{Z(0,\vec k)} &=& Z_{3T} \,+\, \Pi_Z(0,\vec k) \,+\,
                     \sum_{q_0\neq 0}\Pi_Z(q_0,\vec k)\label{eq:geq} \\
\frac{1}{H(0,\vec k)} &=& Z_{3L} \,+\, \sum_{q_0\neq 0}\Pi_H(q_0,\vec k) \; ,
\label{eq:HofZ}
\eea
where we indicate with $\Pi_i$ the various self-energies.
Note that, in the 3d-longitudinal equation the zero-component $q_0=0$
explicitly vanishes, since $P(0,0,\vec k,\vec q)=0$. This is a direct
consequence of the tensor-structure of the ghost-gluon vertex. Thus, the behavior
of the longitudinal mode $H(0,\vec k)$ is entirely determined by the hard modes
and depends only implicitly on the soft ones.

The treatment of the full system, given in Appendix \ref{safull},
justifies a-posteriori this ghost-loop-only truncation for momenta $k$
in the far IR limit. The same truncation is also sensible at zero temperature
\cite{Fischer:2006vf}.
Furthermore, the ghost-gluon vertex is taken equal to the bare one. 
This approximation is supported by various calculations at zero
\cite{Schleifenbaum:2004id,Alkofer,Ilgenfritz:2006gp,Maas:2006qw} and at infinite temperature
\cite{Cucchieri:2006tf,Schleifenbaum:2004id}.
Of course, when considering the far IR limit at non-zero temperature one needs
to ensure the condition $|\vec k|\ll T$ for the external momenta. In this limit
one obtains \cite{Maas:2005hs} that the hard-mode dressing functions
$Z(k_0,\vec k)$, $H(k_0,\vec k)$, $G(k_0,\vec k)$ become constant in the infrared, i.e.\ they
have the behavior of the dressing function of {\em massive} particles. Thus, for
$\vec k \to 0$, the dressing functions of the hard modes for the
3d-transverse gluon, the 3d-longitudinal gluon and the ghost, 
are respectively given by $A_z(k_0)$,
$A_h(k_0)$ and $A_g(k_0)$,
which depend only on the Matsubara mode $k_0$.
In order to find a solution to the system of equations \prefr{geq}{heq}, we
need these IR constants to be bounded.

Finally, for the zero-modes ($k_0 = 0$) we consider the power-law ans\"atze
\bea
G(0,\vec k) &=& B_g\, k^{2\kappa} \label{iransatz1} \\
Z(0,\vec k) &=& B_z\, k^{2t} \\
H(0,\vec k) &=& B_h\, k^{2l} \; ,
\label{iransatz3}
\eea
which depends on the IR exponents $\kappa$, $t$ and $l$ and on the constant
coefficients $B_g$, $B_z$ and $B_h$. Then, for all values of $\kappa$ and $t$,
the self-energies $\Pi_G(0,\vec k)$ and $\Pi_Z(0,\vec k)$ take exactly the same form as
in three dimensions \cite{Maas:2004se}, i.e.\
\begin{widetext}
\bea
\Pi_G(0,\vec k) &=& -B_g\,B_z\, g^2TC_A \frac{2^{1-4\kappa}\Gamma(2+2\kappa)}{
   \kappa(3+4(-2+\kappa)\kappa)\Gamma\left(2\kappa+\frac{3}{2}\right)}y^{t+\kappa-\frac{1}{2}} \\[2mm]
\Pi_Z(0,\vec k) &=& B_g^2\, g^2TC_A \frac{2^{-4(1+\kappa)}(-2+\zeta+2\kappa(\zeta-1))
           \Gamma(2+2\kappa)\sec(2\pi\kappa)\sin(\pi\kappa)^2}{
        \kappa^2(1+\kappa)\Gamma\left(2\kappa+\frac{3}{2}\right)}y^{2\kappa-\frac{1}{2}} \; ,
\label{eq:PiZ}
\eea
where $y = k^2 = \vec k^2$. Clearly, if these contributions are not exactly canceled (or
dominated) by the remaining Matsubara sums, they will give the leading
IR behavior of $G(0,\vec k)$ and of $Z(0,\vec k)$. In this case one finds
a 3d-type behavior for all non-zero temperatures $T$. To show that this is
indeed the case we observe that --- by setting the hard-mode dressing
functions equal to the constants $A_g(k_0)$, $A_z(k_0)$ and $A_h(k_0)$ ---
the self-energies in the ghost equation \pref{eq:geq1} can be rewritten as
\be
\Pi_G(q_0\neq 0,\vec k)\,=\,g^2TC_A\left[-
   \left(\frac{A_z(q_0)}{12\pi}\,+\,
  \frac{A_h(q_0)}{96}\right)\frac{1}{|q_0|}+\left(\frac{A_z(q_0)}{80\pi}\,+\,
   \frac{A_h(q_0)}{1920}\right)\frac{\vec k^2}{|q_0|^3}+\vec k^3\pi(q_0,\vec k)\right] \; .
\label{eq:PiGfinal}
\ee
\end{widetext}
Here, the first term is logarithmically divergent and
it can be absorbed in the renormalization constant. At the same time,
the second term is sub-dominant when compared to the 3d-term, for
values of $\kappa$ and $t$ permitted by integral convergence \cite{Maas:2004se}, and is finite
after summation over $q_0$. Finally, the function $\pi(\vec k)$ vanishes
identically as $\vec k\to 0$. Thus, the leading part of the ghost equation
is the same as in the 3d-case, i.e.\ it is given by $\Pi_G(0,\vec k)$,
provided that the quantities $A_z(q_0)$ and $A_h(q_0)$
do not rise strongly as $q_0$ goes to infinity.
Actually, due to asymptotic freedom, these quantities vanish logarithmically
in the ultraviolet limit.

Finally, we should check that a 3d-type behavior is obtained
for all non-zero temperatures $T$ also when considering the
gluon equations \pref{zeq} and \pref{heq}. To this end, we
first discard spurious divergences in these two
equations. These are quadratic divergences, vanishing for $\zeta=3$ in
the 3d-transverse equation \pref{zeq} and similarly in the longitudinal equation \pref{heq}.
One can show that they are artifacts of the truncation scheme considered
\cite{vonSmekal,Maas:2005hs,Maas:2005rf}.  After subtraction of these
divergences, one can write the contributions of the hard-modes in the self-energies 
[see Eqs.\ \prefr{eq:geq}{eq:HofZ}] as
\begin{widetext}
\be
\Pi_Z^S(q_0,\vec k) \;=\; \frac{\pi_Z(q_0,\vec k)}{k} \; ,  \qquad \qquad
\Pi_H^S(q_0,\vec k) \;=\; \frac{\pi_H(q_0,\vec k)}{k} \; .\label{subconlong}
\ee
In both cases the function $\pi_i$ vanishes as $\vec k\to 0$, separately for each Matsubara term.
Thus, the subtracted part
cannot contribute to the IR behavior in the 3d-transverse equation. This is the case\footnote{Note
that this also applies to the remaining gluon loops, which are not treated explicitly here.}
also after summing over $q_0$. On the other hand, the IR contributions
from the un-subtracted self-energies cannot in general be neglected, as done when considering
only the contributions \pref{subconlong}. This requires,
of course, a regularization \cite{vonSmekal} of the spurious divergences.\footnote{Note
that, in the 3d-transverse case, divergences appear for each hard mode, while in
the 3d-longitudinal case only the sum over the hard modes is affected by spurious divergences.}
To achieve this, one can replace the approximately constant dressing functions
of the hard modes by $A_g(q_0)(q^2+q_0^2)^{-\tau}$, which are clearly suppressed when
$\tau>0$. This is the prescription commonly used for regularizing the divergences at zero
temperature with $\tau=-\kappa$ \cite{Alkofer:2000wg,Fischer:2006ub,vonSmekal,Zwanziger}.
Then, the integrals can be performed and in the limit $\vec k \to 0$ one finds
\bea
\Pi_Z^D(q_0,\vec k\to 0) &=& -\frac{g^2TC_A}{k^2}\frac{(\zeta-3)\Gamma
            \left(\tau-\frac{1}{2}\right)}{32\pi^{3/2}\Gamma(2+2\tau)}(2\pi T)^{1-4\tau}
                    \sum_{n\neq 0}A_g(n)^2|n|^{1-4\tau} \label{imt} \\
\Pi_H^D(q_0,\vec k\to 0) &=& \frac{g^2TC_A}{k^2}\frac{\Gamma
            \left(2\tau+\frac{1}{2}\right)}{8\pi^{3/2}\Gamma(2+2\tau)}(2\pi T)^{1-4\tau}
                    \sum_{n\neq 0}A_g(n)^2|n|^{1-4\tau} \label{iml} \; ,
\eea
\end{widetext}
where $n=q_0/2\pi T$. These sums diverge for $\tau\le 1/2$, due to the term $q_0^{1-4\tau}$.
For $\tau=1/2$, this exponent becomes equal to $-1$, i.e.\ the sums are logarithmically divergent.
Finally, for $\tau>1/2$, the sums are finite and they can (in principle) be resummed analytically.
For example, for $A_g(n)=(2\pi T)^{2\tau}$ one finds that the sum in $n$ is equal to
$2(2\pi T)^{4\tau}\zeta_R(4\tau-1)$, where $\zeta_R$ is the Riemann-$\zeta$ function.
One should also note that the terms in \prefr{imt}{iml} behave like $1/k^2$, i.e.\ like
a mass term in the IR limit. In the 3d-transverse case --- but not in the
3d-longitudinal case \cite{Maas:2005hs} --- this term may not be renormalized, as
this is not allowed by gauge-invariance. If one does not allow either a divergence
stronger than logarithmic (even though this cannot be excluded by the perturbative
renormalization theorems) or different exponents $\tau$ to the 3d-transverse
and to the 3d-longitudinal case, then the only possibility is to have $\tau>1/2$.
For all such values of $\tau$, the contribution \pref{imt} is sub-leading in the 3d-transverse 
equation \pref{eq:geq}. At the same time, the screening mass in the
3d-longitudinal equation would then be solely due to the regularized contribution \pref{iml}. 
This result is a consequence
of the truncation scheme. In fact, due to the decoupling of the soft modes in the
3d-longitudinal equation [see Eq.\ \pref{eq:HofZ}], the result is dominated by the hard
modes, which are very sensitive to truncation artifacts, since they live on a scale that
is effectively mid-momentum. Thus, the present truncation scheme is not able to yield a
consistent description of the electric screening mass and a determination of its value is not
possible. Nonetheless, the 3d-longitudinal gluon propagator seems to exhibit a screening
mass, and thus gives likely a qualitatively correct description of the physics involved. 
This screening mass would not allow the 3d-longitudinal gluon propagator to modify the
IR behavior of either the 3d-transverse gluon propagator or of the ghost propagator.
Therefore, as shown above, these two propagators should  behave at all non-zero
temperatures exactly as in the three-dimensional case when one considers
momenta much smaller than the temperature $T$ and than $\Lambda_{\mathrm{QCD}}$. It is
then plausible that for momenta in the range $T\ll p\ll \Lambda_{\mathrm{QCD}}$ the gluon
and the ghost propagator would still exhibit a behavior similar to that found in
the four-dimensional case. This is also found in DSE studies in a finite volume \cite{fischer}.

The behaviors described in this section are in agreement with calculations
using renormalization-group techniques in a background-field gauge \cite{Braun:2006jd}. Indeed, when
going from $T=0$ to $T > 0$, also in that case one finds a discontinuous
change in the behavior of the running coupling, i.e.\ in the IR region the running
coupling switches from a four-dimensional to an effectively three-dimensional behavior. Let us
also note that previous studies using DSEs \cite{Gruter:2004bb} had {\em assumed} that the
zero-temperature behavior persists up to the thermodynamic transition. 
Here we have relaxed this hypothesis and eliminated some of the 
approximations of the numerical method employed in \cite{Gruter:2004bb},
showing that
the high-temperature results apply also to non-zero temperatures below the critical
temperature $T_c$, a phenomenon previously interpreted as super-cooling \cite{Maas:2005hs}.


\section{Summary}\label{sscenario}

Let us summarize here the results obtained above for the 4d finite-temperature case.
From our lattice calculations we have that:
\begin{itemize}
\item The 3d-transverse gluon propagator decreases as the
      temperature increases. It also seems to have smaller finite-volume
      effects (in the IR limit) and stronger IR suppression
      at high temperature than at zero temperature.
\item The 3d-longitudinal gluon propagator at small momenta
      increases as the temperature goes from 0 to $T_c$, seems to diverge at 
      $T_c$ and then drops as $T$ becomes larger, apparently reaching a 
      plateau for small $p$. Also, in the IR limit,
      it is less divergent than the propagator of a massless particle, with
      a possible exception near the phase transition.
\item The ghost propagator is nearly temperature-in\-de\-pen\-dent.
\item The smallest eigenvalue of the Faddeev-Popov operator, considered as
      a function of the spatial lattice side $L_s$, goes to zero 
      faster than $1/L_s^2$ as $L_s$ goes to infinity,
      for all temperatures.
\end{itemize}
From the DSEs we see that:
\begin{itemize}
\item The IR behavior of the gluon and ghost propagators
      changes abruptly from zero to any non-zero temperature.
\item For momenta $p \ll (T, \Lambda_{\mathrm{QCD}})$, the ghost and the 3d-transverse
      gluon propagators show the same behavior obtained in the dimensionally reduced theory
      \cite{Maas:2004se,Maas:2005hs}
      and the IR behavior in the spatial sector is in accordance with the
      prediction of the Gribov-Zwanziger scenario \cite{Zwanziger,Gribov}.
\item The 3d-longitudinal gluon is likely to have a dynamical mass at
      any non-zero temperature.
\item These results can be connected continuously to
      perturbation theory at large $p$ (see Appendix \ref{safull})
      and to the solution at zero temperature (see Appendix \ref{avac}).
\end{itemize}

As explained in Section \ref{DSEq} above, the main assumption for these DSE
results is that the electric screening mass is non-zero at all temperatures.
Indeed, in the case of a null screening mass, one can find different
solutions for the DSEs. In particular, the ghost propagator may or
may not be IR divergent, while the 3d-transverse gluon propagator
would still be IR suppressed \cite{Maas:2005rf}. Note that the analytic asymptotic
results from the DSE calculations are valid for momenta $p \ll \min(T,\Lambda_{\mathrm{QCD}})$.
As already stressed in Section \ref{svolume}, this range of momenta is not
accessible with the lattice volumes used here. At intermediate momenta, DSE
results agree qualitatively with the lattice results for all temperatures
considered.

We can try to interpret these results by considering the Gribov-Zwanziger
scenario applied to the 4d $T=0$ case and to the dimensionally reduced theory.
As explained in Section \ref{svolume}, this scenario is supported by lattice
\cite{Cucchieri:2006xi,Cucchieri:2003di,Cucchieri:2006tf,Cucchieri:1997dx,Sternbeck:2005tk,Silva} 
and by DSE calculations \cite{vonSmekal,Zwanziger,Maas:2004se}.
Then, at $T > 0$, due to the compactification of the time direction, the configuration space
near the Gribov horizon should be essentially reduced to that of a
three-(space-time)-dimensional system and, in the infinite-volume limit,
the configurations should lie on the Gribov horizon \cite{Gribov} for all
temperatures and the Gribov-Zwanziger confinement scenario
would apply to the so-called magnetic sector.\footnote{This is the case not only
in the Landau gauge but also in Coulomb gauge \cite{Zwanziger:2006sc,Cucchieri:2006hi}. Note that this
is in line with the fact that, in the spatial subspace, the interactions are qualitatively
unaltered when the temperature is switched on and all purely spatial observables,
such as the spatial string tension \cite{Bali}, are not modified by the transition.}
As a consequence, one can remove the so-called IR problem \cite{Zahed:1999tg,
Maas:2005ym,Linde:1980ts,Blaizot:2001nr} since an infrared suppressed
3d-transverse gluon propagator cancels the perturbative infrared divergences.

Based on the above considerations, our results may be organized
into the following scenario:
\begin{itemize}
\item At non-zero temperature, the time-like momenta are no longer continuous, but
discrete. Due to this gap, the properties of configurations near the
Gribov-horizon are changed. This modifies (probably only quantitatively)
the spectrum of the Faddeev-Popov operator. Thus, the enhancement for small
eigenvalues observed at $T = 0$ \cite{Sternbeck:2005vs} should still be present
and the IR enhancement of the ghost propagator might be reduced but not
eliminated. This is verified by our lattice data.
\item The consequences for the 3d-transverse gluon and the 3d-longitudinal gluon
are different, due to the vector character of the ghost-gluon interaction.
In particular, in the 3d-transverse case, the coupling is not modified in the DSEs
and this propagator is still IR suppressed. Moreover, this suppression is actually
stronger due to the structure of the 3d-transverse gluon-ghost interaction.
This is also suggested from our lattice data.
\item
On the other hand, the coupling of the ghost to the 3d-longitudinal gluon becomes
gapped. Thus, the solution of the DSEs is consistent with these gluons acquiring 
a real (dynamical) mass, being no longer IR suppressed. 
This 3d-longitudinal screening mass is generated solely by the hard modes, 
leading likely to a mass of the order of the temperature, which is the characteristic
scale of the hard-modes. In addition, this is observed in lattice calculations
at very high temperatures \cite{Cucchieri}.
At the same time, this screening mass should, similarly to the hard modes, be 
sensitive to the phase transition. This is partially seen from our lattice data,
since the longitudinal propagator is sensitive to the phase transition and
seems to reach a plateau at small $p$ for $T>T_c$.
\item Finally, it should be noted that the presence of an
electric screening mass does not imply that the 3d-longitudinal gluon is
deconfined or an observable particle. Indeed, a massive particle
does not necessarily have a positive-definite spectral function
\cite{Maas:2004se,Maas:2005hs,Maas:2005rf}. Moreover, the $A_0$-field is
(at the perturbative level) member of a BRST-quartet at any finite temperature
\cite{Das:1997gg}. These considerations would imply that no gluon is part
of the physical spectrum also at any $T > 0$.\footnote{Let us note that
this would be in line with the violation of the Oehme-Zimmermann
super-convergence relation \cite{Oehme:1979ai} for the gluons, which is a
pure ultraviolet effect \cite{Nishijima:1993fq} and thus occurs also at non-zero temperature.}
\end{itemize}

A number of predictions follows from the scenario presented above,
which could in principle be further tested using lattice calculations
at sufficiently large volumes and with sufficiently fine discretization.
The main predictions are:
\begin{itemize}
\item For all temperatures, the 3d-transverse gluon propagator is
IR suppressed, while the ghost propagator is IR enhanced.
\item When $p \ll (T, \Lambda_{\mathrm{QCD}})$ the IR behavior of the 3d-transverse gluon
propagator and of the ghost propagator is quantitatively different
from the behavior observed for $T \ll p \ll \Lambda_{\mathrm{QCD}}$, the latter
one being similar to the behaviors observed at zero temperature with
momenta $p\ll\Lambda_{\mathrm{QCD}}$.
\item The 3d-longitudinal gluon propagator is null at zero momentum
only at $T=0$, while it goes to a finite (non-zero) value
at $p=0$ for all temperatures $T > 0$. This propagator is sensitive to the
phase transition.
\end{itemize}

Of course, it would
be interesting to compare our numerical results to a similar study for the $SU(3)$
group, for which a different kind of thermodynamic transition is expected
\cite{Karsch:2001cy}. Note that first investigations, with and without dynamical 
quarks, have been performed in \cite{Furui:2006py}.


\section{Conclusions}\label{ssum}

We have presented an analysis of gluon and ghost propagators
at finite temperature, using lattice gauge theory and DSEs. Results
using these two approaches seem to be (at least at the
qualitative level) consistent with each other. In particular,
when the temperature is turned on, one finds different effects
for 1) the temporal sector, which includes the soft 3d-longitudinal gluon,
2) the spatial sector, containing the soft ghost and the soft
3d-transverse gluon and 3) the hard sector with the (essentially
perturbative) hard modes. In the spatial sub-sector, the IR behavior
is similar to that of a 3-dimensional theory and the correlation
functions agree with those expected from the Gribov-Zwanziger
confinement scenario. Also, their dependence on the temperature
is smooth. On the contrary, the temporal sector is sensitive
to the thermodynamic transition, being screened and decoupling
at the scale of the temperature. Nevertheless, our results are 
consistent with the possibility that all gluons are confined
at all temperatures. Note that this does not alter the bulk
thermodynamic properties, which still have a Stefan-Boltzmann-like behavior
\cite{Maas:2004se,Maas:2005hs,Zwanziger:2006sc,Zwanziger:2004np}.
These results can be understood by considering the Gribov-Zwanziger
confinement scenario at $T \neq 0$. As a consequence, gluon confinement
is not affected by the thermodynamic transition. This would explain
the presence of non-perturbative effects in the high-temperature phase
and the confining properties of the dimensionally-reduced theory
in the infinite-temperature limit.

One should, however, recall that these results are affected by various
(technical) limitations. Indeed, much larger and finer lattices and
more sophisticated DSE schemes are needed in order to obtain
a complete understanding of the effect of the temperature in the
gluonic sector of Yang-Mills theory.

Finally, let us note that the introduction of dynamical quarks should not modify
the Gribov-Zwanziger scenario and, as a consequence, the scenario
described above. At the same time, the chiral transition should also
not be affected by the IR arguments considered here, as the relevant dynamics
occurs at the scale of $\Lambda_{\mathrm{QCD}}$ (see e.g.\ \cite{Fischer:2006ub}).


\acknowledgments

A.\ M.\ is grateful to Reinhard Alkofer and Jochen Wambach for useful discussions. 
A.\ M. was supported by the DFG under grant number MA 3935/1-1.
A.\ C. and T.\ M. were supported by FAPESP (under grants \# 00/05047-5
and 05/59919-7) and by CNPq. The ROOT framework \cite{Brun:1997pa} has been used
in this project.


\appendix

\section{Asymmetric lattices}\label{saasym}

\begin{figure*}[t]
\includegraphics[width=\linewidth]{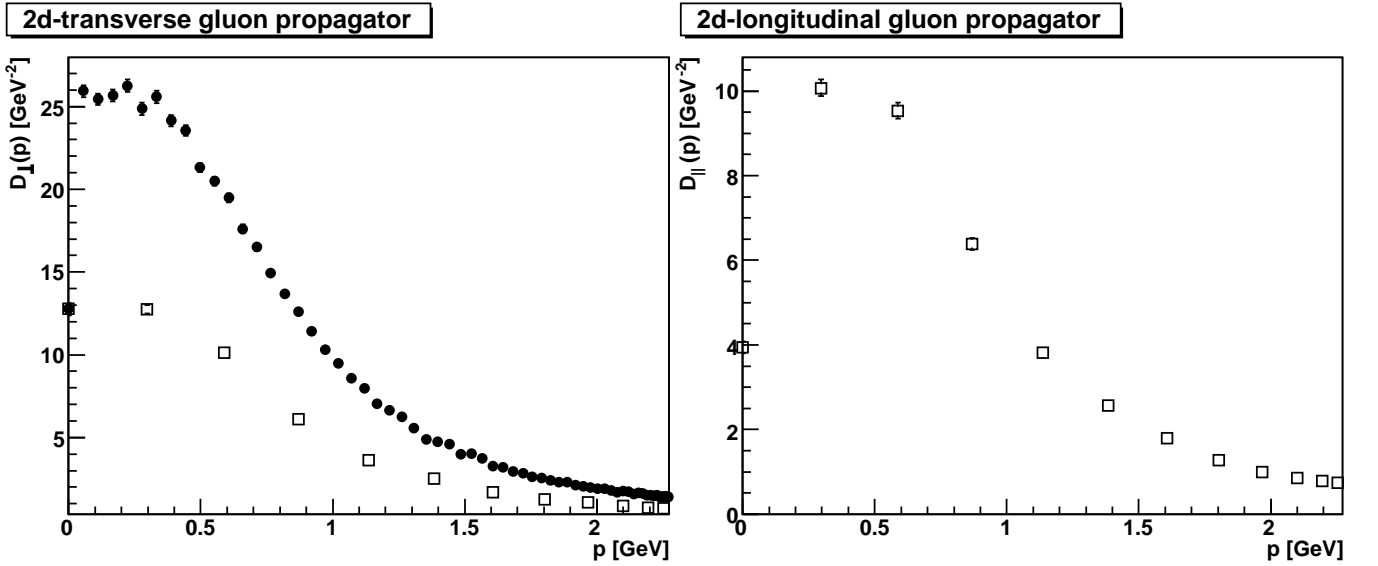}
\caption{\label{asy}
The orthogonal $D_{\perp}(p)$ and the parallel $D_{\parallel}(p)$ components of the gluon propagator
evaluated using the (three-dimensional) strongly-asymmetric lattice $24^2\times 128$ 
at $\beta=4.2$. Open symbols correspond to momenta ($p_s$) along one of the short sides
of the lattice, while full symbols correspond to momenta ($p_l$) along the long side
of the lattice. Here we considered 420 configurations and the same settings used for
{\em Setup} 15. Note that, due to the Landau gauge condition, one has
$D_{\parallel}(p_l,0)=0$ for all momenta $p_l$.}
\end{figure*}

When using asymmetric lattices at $T=0$ (e.g.\ if the time extension of the lattice is
larger than the spatial ones) it is convenient to consider the two scalar gluon
propagators given by $D_{\perp}(p)=D_{ii}(p)/(d-1)$ and $D_{\parallel}(p) = D_{00}(p)$,
where in the first case we sum over the spatial indices $i$ only.
In Fig.\ \ref{asy} we report the results obtained for these two different tensor
components of the gluon propagator when considering an asymmetric lattice of size
$24^2\times 128$ (at $\beta = 4.2$). Let us note that the total volume in this case
is roughly the same
as for a $40^3$ lattice. In this symmetric case, one clearly sees a maximum in the propagator
for a momentum $p \approx 400$ MeV (see Section \ref{svolume} and the top plot in Fig.\ 
\ref{gpv}). In the asymmetric case, on the contrary, it is evident that the two tensor 
components show a very different behavior and
that the behavior of the orthogonal propagator $D_{\perp}(p)$ depends on the type of momentum
considered, i.e.\ along the short axis of the lattice ($p_s$) or along the long
axis of the lattice ($p_l$). Also, there is no clear evidence of a maximum, even
when considering $D_{\perp}(p_l)$. Moreover, as observed already in \cite{Cucchieri:2006za},
the IR suppression obtained using an asymmetric lattice $N_s^{d-1} N_t$ with
$N_t \gg N_s$ is usually smaller than the suppression found on a symmetric lattice
with volume $N_t^d$. This effect clearly reduces the advantages of using asymmetric lattices.

When considering asymmetric lattices, one might also use asymmetric couplings
in order to have the same physical extent in all directions. This would probably
reduce the systematic effects observed in \cite{Cucchieri:2006za}, while keeping
the numerical costs of the simulations small.


\begin{widetext}

\section{Integral kernels}\label{sakernels}

Here we define the integral kernels obtained with the ghost-loop-only truncation
[see Eqs.\ (\ref{geq})--(\ref{heq})]. We use the following notation
\be
\left. \begin{array}{l}
     y=k^2=k_0^2+\vec k^2 \, ; \quad\quad
     x=q^2=q_0^2+\vec q^2 \, ; \quad\quad
     k=|\vec k| \, ;\quad\quad
     q=|\vec q| \\[2mm]
     u=x+y+2k_0q_0-2\vec k \cdot\vec q \, ;\quad\quad
     v=(\vec k-\vec q)^2 \\[2mm]
     z=x+y+2k_0q_0+2\vec k \cdot \vec q \, ;\quad\quad
     w=(\vec k+\vec q)^2 \\[1mm]
     \theta=\arccos{\left(\vec k \cdot \vec q / kq\right)} \; .
       \end{array}
\right. 
\ee
Then, the kernels in the ghost self-energy \pref{geq} are given by
\bea
A_T(k_0,q_0,\vec k,\vec q) &=& \frac{q^2(\vec k \cdot \vec q-k^2q^2) \sin\theta}{xyuv} \,, \\
A_L(k_0,q_0,\vec k,\vec q) &=&-\frac{q^2 \left[ k_0q^2+k^2q_0-(k_0+q_0)\vec k \cdot \vec q
    \right]^2 \sin\theta}{xyvu^2} \; .
\eea
At the same time, in the 3d-transverse gluon equation \pref{zeq}, the kernel is equal to
\be
R(k_0,q_0,\vec k,\vec q) \,=\, 
  -\frac{q^3 \left[ -q+k(\zeta-1)\cos\theta+q\zeta\cos\theta^2 \right] \sin\theta}{2xyz}\,,
\label{kernelr}
\ee
while in the 3d-longitudinal equation \pref{heq} it is
\bea
P(k_0,q_0,\vec k,\vec q) &=& \frac{1}{y^2xz} [q^2 (q_0 \{ -2k_0^2 (k_0 + q0) (\xi - 1)
                               + k^2 [ k_0 - \xi (k_0 - q_0) ] \} \nn \\[2mm]
                               & & \qquad \quad
                        - kq k_0 \{ k_0 (\xi - 1) + 2 q_0 (2 \xi - 1) \} \cos\theta
                               + k_0^2 q^2 \xi \cos\theta^2 ) ] \sin\theta \; .
\label{kernelp}
\eea
Note that the 3d-transverse equation \pref{zeq} has been contracted with a tensor
parameterized by a variable $\zeta$ \cite{Maas:2005hs}. This variable appears only
inside the integration kernel \pref{kernelr}. The case $\zeta=1$ corresponds to the
projector given in Eq.\ \pref{projt}.
The same has also been done for the 3d-longitudinal equation \pref{heq}, with the variable $\xi$.
Again, it only appears inside the integration kernel \pref{kernelp}.
For the zero mode ($k_0=0$), $\xi$ factorizes in all terms and can thus be eliminated from the
equation. Hence the equation depends only implicitly on $\xi$, due to the hard modes \cite{Maas:2005hs}.


\section{Vacuum solution}\label{avac}

As said in Section \ref{DSEq} above, a necessary requirement is that Eqs.\ \prefr{geq}{heq}
reproduce the well-known vacuum behavior \cite{vonSmekal} at $T=0$. To show this, it is
useful to consider linear combinations of Eqs.\ \pref{zeq} and \pref{heq}. In particular,
we consider a combination with weights given by $2/3$ and $1/3$. We also introduce a second combination
with weights equal to $1$ and $(\zeta-3)/2$. At $T=0$ the Matsubara sums become integrals. After performing
these integrals in the equations obtained by considering both linear combinations, we find
\bea
f(k) &=& \frac{2}{3Z}+\frac{1}{3H} = -g^2 B_g ^2\frac{4^{-3-2\kappa}
          \pi^2(1+\kappa)(3+2\kappa)(-6-4\kappa+\zeta(3+4\kappa))
            \csc(2\pi\kappa)}{3\Gamma(1-\kappa)^2\Gamma\left(\kappa+\frac{5}{2}\right)^2}
               y^{2\kappa} \label{sumeq} \\
l(k) &=& \frac{1}{Z}+\frac{\zeta-3}{2}\frac{1}{H} = Z_{3T}+\frac{\zeta-3}{2}Z_{3L}+
                         \frac{\rho(k)}{k^2} \; , \label{diffeq}
\eea
\end{widetext}
with $y=k^2$. The function $\rho(k)$ vanishes for $k\to 0$.
Hence, for $\kappa\ge 0.5$, $l(k)$ is sub-leading compared to $f(k)$. This yields
\bea
Z(k) &=& \frac{2(\zeta-4)}{3(\zeta-3)f(k)-2l(k)}\label{z4d} \\
H(k) &=& \frac{4-\zeta}{3f(k)-2l(k)} \; .
\eea
Without considering the trivial case $\zeta=4$ and neglecting the sub-leading quantity $l(k)$, 
these two functions coincide
only in the case $\zeta=1$. For other values
of $\zeta$ the dressing functions $Z(k)$ and $H(k)$ will have the same leading IR
behavior, but they will differ by a constant factor. This is simply an artifact of the
non-$O(4)$-invariant projection considered. Thus, for the remaining discussion
we can set $\zeta$ equal to 1.

\begin{figure*}
\includegraphics[width=\linewidth]{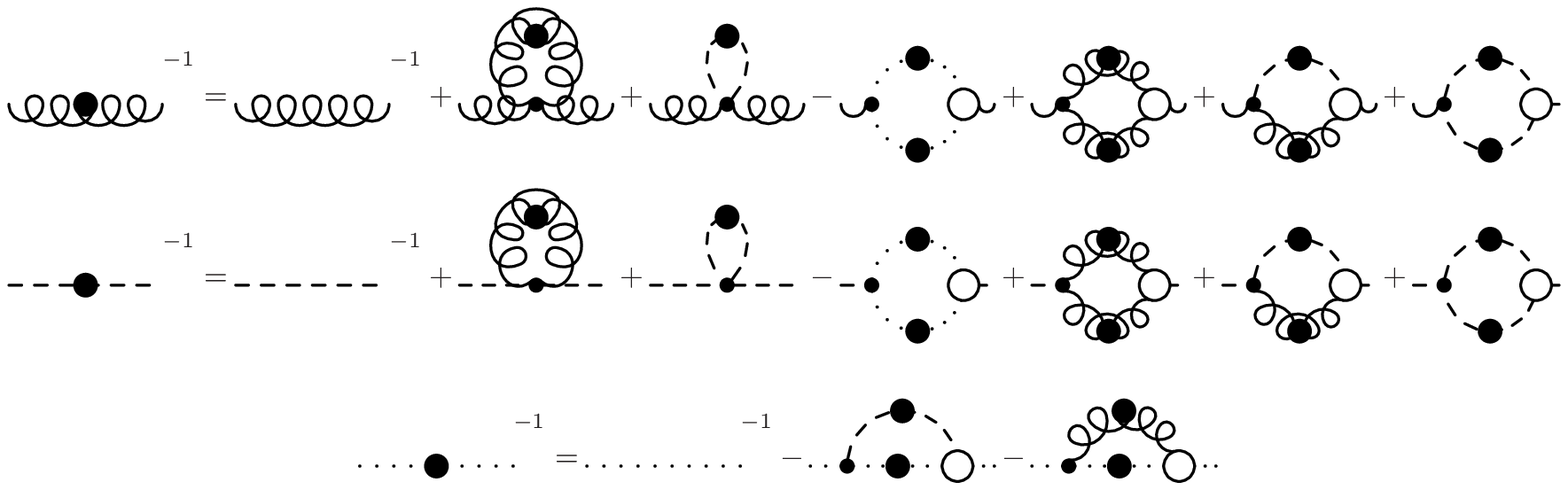}
\caption{The truncated Dyson-Schwinger equations at finite temperature. The wiggly lines
represent transverse gluons, the dashed lines longitudinal gluons and the dotted lines
denote ghosts. Lines with a full dot represent self-consistent propagators while small
dots indicate bare vertices. The open-circled vertices are full and must be constructed in a 
given truncation scheme. A bare ghost-gluon vertex and modified bare gluon vertices
have been used here.}\label{fftsys}
\end{figure*}

By using the above results and the IR ansatz \pref{iransatz1}, the ghost equation
gives the relation
\be
y^{-\kappa} = -g^2B_g^2B_z\frac{3\pi\Gamma(-2\kappa)\Gamma(2+\kappa)
     \csc(\pi\kappa)}{\kappa^2(2-3\kappa+\kappa^2)\Gamma(-\kappa)^3\Gamma(2+2\kappa)}
          y^{\kappa+t} \; .
\label{ge4d}
\ee
Here and henceforth the sub-leading terms proportional to $\rho(k)$ have been dropped. Clearly,
this relation can be satisfied for all values of $y$ (in the infrared limit)
only if the four-dimensional relation \cite{vonSmekal,Watson:2001yv}
\be
2\kappa+t=0 \; 
\ee
is satisfied. By dividing equations \pref{sumeq} and \pref{ge4d}, inserting the
power-law ans\"atze \prefr{iransatz1}{iransatz3} and setting $Z=H$, one finds
a consistency condition for $\kappa$, i.e.\
\be
1=\frac{(2-\kappa)(\kappa-1)}{12(3+4\kappa(2+\kappa))} \; ,
\ee
This equation is identical to the $T=0$ equation and is solved by
$\kappa\approx-0.595353$ \cite{Zwanziger,Lerche:2002ep}.


\section{The full system}\label{safull}

In this section we consider the system of DSEs truncated at the one-loop level
(see Fig.\ \ref{fftsys}). This truncation
has already been discussed in detail in Refs.\ \cite{Maas:2005hs,Maas:2005rf} for the
finite (i.e.\ nonzero) temperature case.


\subsection{Perturbation theory and truncation artifacts}

Many direct experimental verifications of QCD are performed in a regime where perturbation
theory is applicable. Thus, any non-perturbative treatment must make contact with it.
In the case of DSEs (at the presented truncation level) this implies recovering the resummed
leading order perturbation theory \cite{Alkofer:2000wg,Fischer:2006ub,vonSmekal}. Of course,
at finite temperature $T$, the vacuum perturbation theory is expected to be recovered only
at momenta large compared to $T$.

In order to show that this is the case, the first step is to identify and remove the
spurious quadratic divergences \cite{Alkofer:2000wg,Fischer:2006ub,vonSmekal}. Since the
summation over the Matsubara frequencies extends to infinity, spurious divergences might
depend on the order of integration and summation. Furthermore, it is possible that a
contribution is finite with respect to integration but not to summation and vice-versa.
Of course, only the region corresponding to large $n$ is relevant for the analysis of
spurious divergences. In this limit, the difference between two consecutive Matsubara
frequencies becomes negligible and the summation becomes equivalent to an integration.
One can then replace $|\vec q|$ by $q\sin\eta$ and $q_0$ by $q\cos\eta$ (with $q^2=q_0^2+\vec q^2$)
in the integral kernels $K(q)$ appearing in DSEs. Since the angular integration over $\eta$ yields
no divergences, it is sufficient to replace all integral kernels $K(q)$ by the prescription
\be
K(q)\to K(q)-\frac{1}{q^2}\lim_{q\to\infty} q^2 K(q) \; .
\label{subp}
\ee
The resulting integrals have no spurious quadratic divergences, but only
the usual logarithmic one. One could absorb the subtracted part in gauge
non-invariant counter-terms, as they originate purely from violation of
gauge invariance. On the other hand, as discussed in Section \ref{DSEq} above, this
affects the IR behavior of the solutions. To surpass this problem constructively,
the electric screening mass is here explicitly renormalized to a fixed value $m_r=g^2T$,
since this is allowed by gauge invariance \cite{Maas:2005hs}. Note that this prescription
is sub-leading in the ultraviolet and does not affect perturbation theory. At the same time,
the corresponding terms in the 3d-transverse equation can be dropped, as they are sub-leading
both in the IR and in the ultraviolet limit. The only exception is the soft-mode contribution,
which is treated as in the infinite-temperature limit \cite{Maas:2004se}.
As for the genuine tadpoles, they could in principle give a finite contribution.
However, when removing the quadratic divergences according to
\pref{subp}, they vanish identically. Thus, also at finite temperature, the tadpoles 
contribute only a pure divergence, which can be absorbed in gauge non-invariant counter-terms.
Also note that this implies that all spurious divergences are not affected by a
finite temperature.\footnote{This could be the case, for an
inconsistent truncation scheme.} This is to some extent not surprising, as the usual
divergences in quantum field theory may not be affected by a finite temperature \cite{Das:1997gg}.

As a second step, we have now to fix the renormalization prescription for the (physical)
logarithmic divergences. In order to recover the 3d-theory in the infinite-temperature limit, 
two aspects have to be considered in the renormalization prescription \cite{Maas:2005hs}. 
Firstly, one needs a finite 3d-coupling. To this end, one requires the 
quantity $g^2T$ to be fixed to $\Lambda_{\mbox{\footnotesize QCD}}$ for large temperatures 
\cite{Maas:2005hs,Maas:2005rf}.\footnote{The precise value of $\Lambda_{QCD}$ is irrelevant
for the present purpose but, as for the zero temperature case \cite{Alkofer:2000wg}, it could
be obtained from comparison with the running coupling at sufficiently large (perturbative) momenta.}
This implicitly defines a temperature-dependent renormalization scheme and
automatically fixes the renormalization scale $\mu$
as a function of $T$ by the running of the coupling constant $g^2/(4\pi)=\alpha_S$. 
At the same time, a smooth zero-temperature limit is obtained using the prescription
\be
\mu=\Lambda_{\mbox{\footnotesize QCD}}\left\{\exp\left[\frac{\left(\sqrt{8}\pi\right)^2}{\beta_0}\right]-
         1+\exp\left(\frac{8\pi^2 T}{\beta_0\Lambda_{\mbox{\footnotesize QCD}}}\right)\right\} \; .
\ee
Here $\beta_0$ is the first coefficient of the $\beta$-function. Then, according to the
perturbatively resummed 1-loop result, the largest value attained by $\alpha_S$ is $1$, i.e.\
$g^2=4\pi$.

Secondly, the wave function must be renormalized such that the dressing functions become
unity at infinite momentum for $T\to\infty$. This is guaranteed by selecting the subtraction point
\be
s=10^4\Lambda_{\mbox{\footnotesize QCD}}+2\pi T \; .
\ee
Note that the term $10^4\Lambda_{\mbox{\footnotesize QCD}}$ has been added to
renormalize perturbatively at zero temperature. Clearly, the choice of temperature-dependent
values for $s$ and $\mu$ entails that the renormalization constants will also contain
finite temperature-dependent contributions.

The regularization and renormalization are performed as in 
Refs.\ \cite{Maas:2005hs,Maas:2005rf} by adding explicit
counter-terms. This makes multiplicative renormalizability manifest.
The renormalization conditions are
\be
G(s)=Z(s)=H(s)=1
\label{rencond} \; .
\ee
They implement the requirement 
\be
G^2(s)Z(s)=G^2(s)H(s)=1
\ee
from Slavnov-Taylor identities (STI) at zero temperature \cite{vonSmekal}.
Note that this prescription does not allow any freedom in the values
of the IR pre-factors in the ans\"atze \prefr{iransatz1}{iransatz3}.
This renormalization scheme is explicitly independent of any Matsubara frequencies, and thus
it suffices to determine the wave function renormalization constants in the soft equations.
On the other hand, this renormalization scheme only works if performed at
momenta sufficiently far in the perturbative regime, since $g$ has only been included to
the leading perturbative order.

In order to recover perturbation theory, the ultraviolet asymptotic dressing functions
must be given by the renormalization-group-improved perturbative propagators \cite{Yndurain:1999ui}
\bea
G(p)&=&G(s)\left[\omega\log\left(\frac{p^2}{s^2}\right)+1\right]^\delta\label{gpp} \\
Z(p)=H(p)&=&Z(s)\left[\omega\log\left(\frac{p^2}{s^2}\right)+1\right]^\gamma\label{zpp} \\
\omega&=&-\frac{3g^2(\mu)C_A}{64\pi^2\delta}\label{omega} \\
\delta&=&-\frac{1}{2}(\gamma+1)=-\frac{9}{44}\label{dgrel} \; .
\eea
This also implies that the longitudinal and transverse gluon propagators must coincide 
at sufficiently large momenta $p$. At the same time, any kind of gauge-invariance violation
must vanish, i.e.\ the propagators should become independent of the variables $\zeta$ and $\xi$.

It thus remains to test whether the truncation scheme permits this. At $T = 0$
it has been found that only a dressed three-gluon vertex yields the correct perturbative
solution \cite{vonSmekal}. Thus, we consider the Bose-symmetric ansatz
\begin{widetext}
\bea
\Gamma(p,q,k)_{\mu\nu\rho} & = & A(q,p,k) \, \Gamma(p,q,k)^{(\mathrm{tl})}_{\mu\nu\rho}(p,q,k) \\
  & = & a \; G\left(\frac{1}{2}\left(q^2+p^2+k^2\right)\right)^{a_G}
        Z\left(\frac{1}{2}\left(q^2+p^2+k^2\right)\right)^{a_Z}
        \Gamma(p,q,k)^{(\mathrm{tl})}_{\mu\nu\rho}(p,q,k) \; .
\label{g3v}
\eea
Here, $\Gamma$ and $\Gamma^{(\mathrm{tl})}$ are respectively the full and the tree-level
vertex and the constants $a$, $a_G$ and $a_Z$
will be chosen such that the propagators \prefr{gpp}{zpp} are a solution of the system.
Of course the above ansatz is valid only at high momenta. The low-momentum case, for which
$Z\neq H$, will be discussed below.\footnote{As usual, due to the STI
and the results obtained in the non-perturbative regime \cite{Alkofer}, the ghost-gluon
vertex is kept bare.}

Finally, it is sufficient to inspect the limit $p\gg p_0$ in a given equation, 
which thus reduces to the corresponding soft equation. As only the large-momentum behavior
is relevant, the integrals can be treated with a lower cutoff $p$. Furthermore, the
denominators can then be expanded in inverse powers 
of the integration momentum\footnote{Originally, at $T=0$, the angular integrals were
solved exactly instead of considering this expansion \cite{vonSmekal}. However, due
to the non-Euclidean invariant projection, this is not as simple anymore.}
up to ${\cal O}(q^{-3})$.
Approximating further internal dressing functions by $G(p+q)\approx G(q)$ etc.\ 
\cite{vonSmekal}, a simple set of equations is obtained
\bea
\frac{1}{G(p^2)}&=&\tilde Z_3-\frac{3g^2C_A}{64\pi^2}\int_{p^2}^{\Omega^2}dy
                                  \frac{G(y)Z(y)}{y} \label{geqpt} \\
\frac{1}{Z(p^2)}&=&Z_{3T}+\frac{g^2C_A(3\zeta-5)}{384\pi^2}\int_{p^2}^{\Omega^2}dy
                               \frac{G(y)^2}{y}-\frac{g^2C_A(47+3\zeta)}{384\pi^2}
                               \int_{p^2}^{\Omega^2}dy\frac{A(y,y,p^2)Z(y)^2}{y} \label{zeqpt} \\
\frac{\xi}{Z(p^2)}&=&\xi Z_{3L}-\frac{g^2C_A\xi}{192\pi^2}\int_{p^2}^{\Omega^2}dy
                               \frac{G(y)^2}{y}-\frac{25g^2C_A\xi}{192\pi^2}
                               \int_{p^2}^{\Omega^2}dy\frac{A(y,y,p^2)Z(y)^2}{y} \label{heqpt} \; .
\eea
\end{widetext}
(The integral kernels and other details of these equations can be found in \cite{Maas:2005rf}.)
In these equations, the asymptotic form \pref{zpp} has already been used to replace $H$ by 
$Z$ and $y=q^2=q_0^2+\vec q^2$. Note that the third equation is the original 3d-longitudinal equation 
and the $\xi$-dependence can obviously be removed. The wave function renormalization constants
are chosen to cancel exactly the upper boundary of the integrals, thus rendering the equations
finite and independent of the regulator $\Omega$.

\begin{figure}
\includegraphics[width=0.91\linewidth]{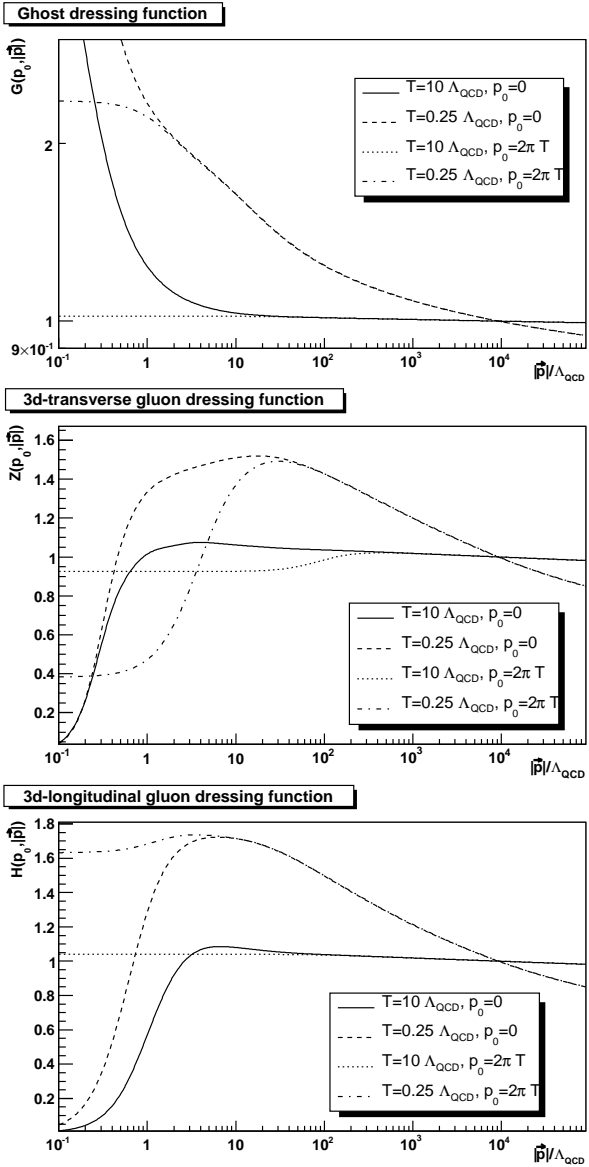}
\caption{\label{ndse}Numerical results from the Dyson-Schwinger equations
at two temperatures in units of $\Lambda_{\mbox{\footnotesize QCD}}$ for the soft mode
$p_0=0$ and for the first hard mode $p_0 = 2 \pi T$ (4 and 5 modes have been calculated
explicitly in the high and in the low temperature case, respectively). We show in the top,
middle and bottom panel the dressing functions of the ghost, of the 3d-transverse gluon
and of the 3d-longitudinal gluon, respectively. The different ultraviolet behavior for different
temperatures is due to the temperature-dependent choice of the
renormalization conditions.}
\end{figure}

One can then check that the ghost equation \pref{geqpt} is solved by \prefr{gpp}{zpp}.
This yields the relation \pref{dgrel} between $\gamma$ and $\delta$ and the value of
$\omega$ [see Eq.\ \pref{omega}]. 
On the other hand, the value of $\delta$ is not fixed. The remaining task
is the choice for $A(y,y,p^2)$. Using the ansatz \pref{g3v} in the 3d-transverse equation \pref{zeqpt},
one finds for the parameters $a_Z$ and $a$ the conditions
\bea
a_Z&=&\frac{-2-6\delta+a_G\delta}{1+2\delta}\label{azcond} \\
a&=&\frac{-18-41\delta+3\delta\zeta}{\delta(47+3\zeta)} \; .
\eea
At the same time, the equation becomes independent of $a_G$.
In the 3d-longitudinal equation \pref{heqpt} one finds again for $a_Z$ the solution \pref{azcond}.
On the other hand, $a$ is determined to be given by
\be
a=\frac{-9-19\delta}{25\delta} \; .
\ee
Incidentally, there is only one value of $\delta$ that at the same time removes the 
$\zeta$-dependence in the 3d-transverse equation and yields the same $a$ value in both 
the 3d-transverse equation and in the 3d-longitudinal equation. This value is $\delta=-9/44$, 
which coincides with the result from the renormalization-group-improved perturbation theory 
\pref{dgrel}. For this value, one finds $a=1$ in both equations above. Thus, the condition of
gauge invariance uniquely requires for this truncation scheme the correct value of $\delta$. 
This is quite convenient, although likely accidental.

Let us note that it is always possible to rewrite $a=Z_1 b$ with an arbitrary value for $Z_1$, thus 
realizing the STI $Z_1=Z_3/\tilde Z_3$ deliberately by fixing $b$ appropriately.

Finally, as said above, we can consider the low-momentum form of the three-gluon vertex.
To this end, we replace $G^{a_G}Z^{a_Z}$ in Eq.\ \pref{g3v} with the quantity
$G^{na_{GT}/3+ma_{GL}/3}Z^{na_Z/3}H^{ma_H/3}$. Here, $n$ and $m$ are the number
of 3d-transverse and 3d-longitudinal legs. The exponents $a_Z$ and $a_H$ are
chosen to satisfy \pref{azcond} for $a_G=a_{GT}$ and $a_G=a_{GL}$, respectively.
A convenient choice for the remaining parameters $a_{GT}$ and $a_{GL}$ is
\bea
a_{GT}&=&\frac{2t(1+3\delta)}{t\delta+\kappa+2\delta\kappa} \\
a_{GL}&=&\frac{2l(1+3\delta)}{l\delta+\kappa+2\delta\kappa} \; ,
\eea
where $\kappa$, $t$, and $l$ are the IR exponents of the dressing functions 
\prefr{iransatz1}{iransatz3}. This yields an IR (positive) constant three-gluon vertex. 
Although this is not in agreement with recent results on the IR behavior of this vertex 
\cite{Alkofer,Maas:2006qw,vrtx4d}, this is irrelevant, because the gluon loops are still
IR sub-leading. 

It remains to select for the numerical solution at intermediate momenta
the Matsubara frequency for which the dressing functions should be 
evaluated in the function $A$. The quantity $\sqrt{q_0^2+p_0^2+(q_0+p_0)^2}/(4\pi T)$ is, 
apart from $p_0=q_0=0$ (the most relevant case), in general not an integer. As there is 
no obvious possibility, the largest integer smaller than this expression will be taken 
as the Matsubara frequency at which the expression is evaluated.

Note that in the present approach the use of the Matsubara formalism and of the 
vacuum perturbation theory in an asymptotically-free theory does not contradict the 
Narnhofer-Thirring theorem \cite{Narnhofer:1983hp}. In fact, solving the set of DSEs
self-consistently is not equivalent to expanding around a free system and the obtained
propagators thus describe thermal quasi-particles.


\subsection{Numerical solution}

A full numerical solution of the DSEs (for an explicit form of the equations see \cite{Maas:2005hs})
and the present vertex construction can be performed using the method described 
in \cite{Maas:2005xh}. Let us note that, in order to improve the speed of the algorithm,
it is useful to let the damping constants, introduced in Ref.\ \cite{Maas:2005xh}, decrease
with the iteration number, starting from a very large value.
In practice, however, only a (small) finite set of the Matsubara frequencies can be treated 
independently, due to limitations in computing power. For the Matsubara frequencies not
determined explicitly we use the perturbative behavior reported in Eqs.\ \prefr{gpp}{zpp}. 
Thus, at sufficiently large momenta, perturbation theory becomes dominant and the
$O(4)$-invariance is restored. We also approximate the summation as an integration,
starting from a given frequency (larger than the largest one determined independently). 
This integration extends to negative and positive infinity, respectively, and can be treated
using a normal Gauss-Legendre integration.

Results for two different temperatures are shown in Fig.\ \ref{ndse}
for the $SU(3)$ case, with $\xi=1$ and $\zeta=3$ (i.e.\ $\kappa=-1/2$)
\cite{Maas:2004se}. These results are qualitatively similar to our findings
using lattice calculations (see Section \ref{slattice}) and to results reported in
Refs.\ \cite{Cucchieri}. A further interesting observation is the temperature dependence
of the IR coefficients, e.g.\ the one of the ghost propagator ($a_G$) increases with
decreasing temperature.

Note that the temperatures considered here are quite large. While
there is no problem in going to arbitrarily large temperatures and obtaining
the infinite temperature limit explicitly, going to smaller temperatures is limited
by two technical problems. One is the sheer computing power and the necessity to include
more Matsubara frequencies. The other one is a truncation-induced problem. Indeed, the
mid-momentum behavior of the three-gluon vertex destabilizes the
system, as the only possible solution would have a sign change in
the gluon propagator, which is not allowed due to Eqs.\ \prefr{dt}{dl}. 
In order to reduce this problem, the numerical calculations have been done using
an interpolation between the perturbative and the IR behavior, instead of considering the 
full dependence on the dressing functions. Such a change does not induce an error
larger than the one already present due to the truncation scheme.
On the other hand, this artifact can only be cured by an adequately chosen effective
vertex-dressing with a stronger mid-momentum suppression. This is actually what is expected
from recent lattice calculations \cite{Cucchieri:2006tf,Maas:2006qw,vrtx4d}.
The same problem appears when considering $\zeta\neq 3$. Thus, these results 
should be taken to be a proof-of-principle that this system indeed has solutions of 
the described type, rather than a quantitative investigation.


\end{document}